\documentclass[onecolumn,12pt,journal,final]{IEEEtran} %arXiv
\def\ps@headings{%
\def\@oddhead{\mbox{}\scriptsize\rightmark \hfil \thepage}%
\def\@evenhead{\scriptsize\thepage \hfil \leftmark\mbox{}}%
\def\@oddfoot{}%
\def\@evenfoot{}}
\makeatother 

\IEEEoverridecommandlockouts
\usepackage{bbm}
\usepackage{amsfonts}
\usepackage[dvips]{graphicx}
\usepackage{times}
\usepackage{cite}
\usepackage{amsmath}
\usepackage{array}
\usepackage{amssymb}

\usepackage{stfloats}
\usepackage{graphicx}
\usepackage{footnote}
\usepackage{booktabs}
\usepackage{array}
\usepackage{algorithmic}
\usepackage{algorithm}
\usepackage{subeqnarray}
\usepackage{cases}
\usepackage{threeparttable}
\usepackage{color}
\usepackage{hyperref}
\usepackage{epstopdf}
\usepackage{cleveref}
\usepackage{bm}
\usepackage{dsfont}
\usepackage[normalem]{ulem} % \delete

\begin{document}
\title{User-Centric Joint Access-Backhaul Design for Full-Duplex Self-Backhauled Wireless Networks}
% \title{Joint Access-Backhaul Design with User-Centric Clustering and Beamforming for Self-Backhauled 5G Networks}
\author{Erkai Chen, Meixia Tao, and Nan Zhang \\
% \author{Erkai~Chen,~\IEEEmembership{Student~Member,~IEEE,} and Meixia~Tao,~\IEEEmembership{Fellow,~IEEE} \\
% \thanks{This work is supported by the National Natural Science Foundation of China under grants 61571299 and 61521062.}
\thanks{This paper was presented in part at IEEE GLOBECOM 2018 \cite{Chen_GLOBECOM18}.}
\thanks{E. Chen and M. Tao are with the Department of Electronic Engineering, Shanghai Jiao Tong University, Shanghai 200240, China (email: cek1006@sjtu.edu.cn; mxtao@sjtu.edu.cn).}
\thanks{N. Zhang is with the Wireless Algorithm Department, Product Research and Development System, ZTE Corporation, Shanghai 201203, China (e-mail: zhang.nan152@zte.com.cn).}
}
\maketitle

\begin{abstract} 
Full-duplex self-backhauling is promising to provide cost-effective and flexible backhaul connectivity for ultra-dense wireless networks, but also poses a great challenge to resource management between the access and backhaul links. In this paper, we propose a user-centric joint access-backhaul transmission framework for full-duplex self-backhauled wireless networks. In the access link, user-centric clustering is adopted so that each user is cooperatively served by multiple small base stations (SBSs). In the backhaul link, user-centric multicast transmission is proposed so that each user's message is treated as a common message and multicast to its serving SBS cluster. We first formulate an optimization problem to maximize the network weighted sum rate through joint access-backhaul beamforming and SBS clustering when global channel state information (CSI) is available. This problem is efficiently solved via the successive lower-bound maximization approach with a novel approximate objective function and the iterative link removal technique. We then extend the study to the stochastic joint access-backhaul beamforming optimization with partial CSI. Simulation results demonstrate the effectiveness of the proposed algorithms for both full CSI and partial CSI scenarios. They also show that the transmission design with partial CSI can greatly reduce the CSI overhead with little performance degradation.
\end{abstract}


\section{Introduction}
Backhaul has emerged as a new challenge of 5G networks due to the ultra-dense deployment of small cells and the explosive growth of data traffic \cite{Yuan_5G_ZTE15,Ge_wireless_backhaul_network14}. 
Traditional fiber-based backhaul can provide high data rates, but the prohibitive cost and the geographical limitations make it impossible to deploy in many practical scenarios. In-band wireless backhauling, also referred to as self-backhauling, is a promising and viable alternative since it utilizes the same spectrum and the same infrastructure with the access link and enables low-cost and plug-and-play installation \cite{Ge_wireless_backhaul_network14}. 
Using self-backhauling, a small base station (SBS) can easily receive data from a macro base station (MBS) in the downlink (or a mobile user in the uplink) and then transmit the data to a mobile user (or an MBS) over the same wireless radio spectrum. Combining recent advances in in-band full-duplex (IBFD) technique further facilitates self-backhauled SBSs to transmit and receive at the same time, thereby potentially doubling the spectrum efficiency \cite{Pitaval_MWC15_self_backhauling,Shojaeifard_FD_CRAN_TWC18}.

Radio resource management across the access and backhaul links is crucial for performance optimization in self-backhauled wireless networks. It is also an important study issue mentioned in the 3GPP technical report on integrated access and backhaul \cite{3GPP_IAB_2018}. The challenge lies in the newly introduced cross-tier interference by spectrum sharing between the access and backhaul links. 
Recently, several research efforts have been made to address the resource management issue for full-duplex self-backhauled wireless networks \cite{Sharma_TWC17_selfbackhauling,Tabassum_TCOM16_massiveMIMO_FD,Chen_JSAC16_IBFD_massiveMIMO,Debbah_EWC16_IBFD_massiveMIMO}. 
The authors in \cite{Sharma_TWC17_selfbackhauling} demonstrated that compared with a conventional time-division duplex (TDD)/frequency-division duplex (FDD) self-backhauled network, the downlink rate in the full-duplex self-backhauled heterogeneous network is nearly doubled, but at the expense of reduced coverage due to higher interference. 
By employing massive MIMO and transmit beamforming at the MBS, the authors in \cite{Tabassum_TCOM16_massiveMIMO_FD} derived the downlink coverage probability of a small cell user considering both the in-band and out-band full-duplex modes of a given SBS. 
Besides, the authors in \cite{Chen_JSAC16_IBFD_massiveMIMO} studied a joint cell association and power allocation problem for energy efficiency maximization. 
An advanced block digitalization precoding scheme was proposed to eliminate the cross-tier interference and multi-user interference.
The authors in \cite{Debbah_EWC16_IBFD_massiveMIMO} studied a joint scheduling and interference mitigation problem for network utility maximization. 
A regularized zero-forcing precoding scheme and the SBS operation mode switching (between full-duplex and half-duplex) were jointly considered to mitigate both the co-tier interference and the cross-tier interference.
Note that in \cite{Sharma_TWC17_selfbackhauling,Tabassum_TCOM16_massiveMIMO_FD,Chen_JSAC16_IBFD_massiveMIMO,Debbah_EWC16_IBFD_massiveMIMO}, it is assumed that each user is associated with only one SBS in the access link and no inter-site cooperation is considered.

Multi-cell cooperation (e.g., CoMP) is a promising technique to mitigate the inter-cell interference by allowing the user data to be jointly processed by several interfering cell sites, 
thus mimicking a large virtual MIMO system \cite{cooperative_jsac,Li_interference_ZTE15}.
Exploiting the inter-site cooperation, the authors in \cite{Zhuang_ICC17_CoMP} studied the joint access and backhaul resource management for ultra-dense networks. Both joint transmission (CoMP-JT) and coordinated beamforming (CoMP-CB) are considered in the access link. However, the cross-tier interference between the access and backhaul links is ignored.
The authors in \cite{Hu_TWC17_multicast_clustering} studied the joint access and backhaul design for the downlink of cloud radio access networks (C-RANs). 
Therein, a user-centric clustering strategy is assumed in the access link such that each user can be served cooperatively by a cluster of nearby SBSs, and multicast transmission is adopted in the backhaul link by the MBS to deliver the message of each user to the SBSs in its serving cluster simultaneously. The messages of different users are delivered one by one in a time-division manner over the backhaul link. 
Based on a similar network model in \cite{Hu_TWC17_multicast_clustering}, a different transmission scheme is considered in \cite{Hua_TWC18} to balance the tradeoff between the cooperation benefits and the coordination overhead, where CoMP-CB is adopted in the access link such that each user is served by only one SBS and accordingly, multi-user beamforming is adopted in the backhaul link. Both works, however, are for out-band wireless backhaul and thus did not have the cross-tier interference issue.

In this paper, we propose a joint access-backhaul transmission framework for full-duplex self-backhauled wireless networks by considering the inter-site cooperation.
As in \cite{Hu_TWC17_multicast_clustering}, an adaptive user-centric clustering strategy is adopted in the access link, where each user is cooperatively served by a cluster of SBSs. 
To reduce the backhaul traffic load, a user-centric multicast transmission is adopted in the backhaul link, where each user's data is treated as a multicast message and the SBS cluster receiving the same user's message forms a multicast group. 
Different from \cite{Hu_TWC17_multicast_clustering}, we consider IBFD-enabled SBSs which can transmit in the access link while receiving over the backhaul link simultaneously using the same frequency band with potential cross-tier interference as well as self-interference. 
In addition, unlike \cite{Hu_TWC17_multicast_clustering} where all users' messages are fetched in a time-division manner over the backhaul link, we adopt a non-orthogonal multicast transmission scheme over the backhaul link so that the MBS delivers all the multicast messages simultaneously, which is more efficient than the orthogonal time-division manner \cite{Ding_NOMA_CM17}.
We also note that the considered system model is similar to full-duplex relay systems with multiple relay nodes \cite{Kim_IBFD_survey2015}, where the MBS can be viewed as the source node, the SBSs can be viewed as the relay nodes, and the users can be viewed as the destination nodes. However, unlike the existing works on full-duplex relay systems, our work differs fundamentally in the problem formulation and the transmission model. For example, the authors in \cite{Krikidis_TWC12} considered a cooperative full-duplex relay system and studied the relay selection problem to achieve the selection diversity gain by choosing the proper relay nodes. However, only one source-destination pair is considered in \cite{Krikidis_TWC12}. The authors in \cite{Ng_TCOM12} considered an MIMO-OFDMA full-duplex relay system with one source, multiple relays, and multiple destinations. A joint user scheduling and resource allocation problem is studied to maximize the system throughput. However, the coverage of the considered network is divided into multiple non-overlapped areas, each corresponding to one relay. There is no cooperation among the relays. By contrast, we consider a more general model with multiple relays cooperatively serving multiple destinations. Each destination can be cooperatively served by a cluster of relays, and the relay clusters of the destinations can overlap with each other. In addition, we adopt multicast transmission to deliver the destinations' messages to their corresponding serving relays. Under this transmission model, we study the joint relay clustering and source/relay beamforming design to seek the maximum achievable rate of the network. 

Enabling inter-site cooperation in general requires acquiring channel state information (CSI) and exchanging it among cell sites, which is a challenging issue for ultra-dense networks. With a large number of cell sites and mobile users in the network, acquiring full CSI will bring excessive signaling overhead and counteract the performance gains provided by cooperative transmission \cite{Wang_WC15,Pan_CM18}.
Specifically, obtaining the CSI of all cell sites at the user receivers through downlink pilot training may require a long training period that is comparable to the channel coherence time. Moreover, feeding back all the CSI by the user receivers will occupy plenty of the uplink resources \cite{Shi_ICC14}. 
After collecting the CSI, each cell site will forward it to the central controller or share it among the cooperating cell sites for joint signal processing. The overhead will easily overwhelm the backhaul resources, especially in self-backhauled wireless networks, where the backhaul is a scarce resource. 
Thus, in this work we consider the user-centric joint access-backhaul design with both full CSI and partial CSI. While the former with full CSI can provide the performance upper bound in the ideal case, the latter with partial CSI offers a more practical solution. The contributions of this paper are summarized as follows.

We first consider the joint access-backhaul design with full CSI, where the CSI between each user and each cell site is globally available at a central controller. We formulate an optimization problem for the joint design of multicast beamforming in the backhaul link, SBS clustering and beamforming in the access link to maximize the weighted sum rate of all users under per-BS peak power constraints. 
This problem is a non-convex mixed-integer non-linear programming (MINLP) problem, which is challenging due to the non-smoothness and the non-convexity of the objective function as well as the combinatorial nature of the SBS clustering.
To solve the problem, we first consider the joint access-backhaul beamforming design problem under given SBS clustering. Note that this problem can be approximately transformed into a manifold optimization problem and solved via the Riemannian conjugate gradient (RCG) algorithm \cite{Absil_manifold_optimization}, as shown in our prior conference paper \cite{Chen_GLOBECOM18}. However, the RCG algorithm may get stuck in unfavorable local points when the MBS peak power is too large. 
In this paper, we propose to solve the joint access-backhaul beamforming problem via the successive lower-bound maximization (SLBM) approach. A novel concave lower-bound approximation for the achievable rate expression in the objective function is introduced based on signal-to-interference-plus-noise ratio (SINR) convexification.
Simulation results show that the proposed SLBM algorithm with the newly introduced SINR-convexification based lower-bound approximation can avoid the high-power issue in the RCG algorithm. 
It can also achieve better performance than the well-known weighted minimum-mean-square-error (WMMSE) algorithm.
We then develop a heuristic algorithm to determine the SBS clustering based on the iterative link removal technique. The effectiveness of the proposed clustering algorithm is also demonstrated via numerical simulations.

We also consider the joint access-backhaul beamforming design with partial CSI, where only part of the CSI in the access link is available. We formulate a stochastic beamforming design problem to maximize the average weighted sum rate of the network under the per-BS power constraints. We develop a stochastic SLBM algorithm to solve it by adopting the introduced SINR-convexification based lower-bound approximation. Moreover, we derive a deterministic lower-bound approximation for the average achievable rate by using Jensen's inequality. The original stochastic optimization problem is then approximately solved by solving the resulting deterministic approximation with low complexity. 
Simulation results demonstrate the performance of the proposed stochastic SLBM algorithm as well as the effectiveness of the proposed deterministic lower-bound approximation. The results also indicate that with a moderate amount of CSI, the proposed algorithms can achieve good performance that is very close to the full CSI case and significantly reduce the channel estimation overhead. 

% \subsection{Organization and Notations}
The rest of the paper is organized as follows. Section \ref{sec:System Model} introduces the system model. Section \ref{sec:Joint Access-Backhaul Design with Full CSI} considers the joint access-backhaul design with full CSI and introduces the proposed joint access-backhaul beamforming and SBS clustering algorithms. The joint access-backhaul design with partial CSI is presented in Section \ref{sec:Joint Access-Backhaul Design with Partial CSI}. Simulation results are provided in Section \ref{sec:Simulation Results}. Finally, we conclude the paper in Section \ref{sec:Conclusion}.

\textit{Notations}: 
Boldface lower-case and upper-case letters denote vectors and matrices, respectively. Calligraphy letters denote sets or problems, depending on the context. $\mathbb{R}$ and $\mathbb{C}$ denote the real and complex domains, respectively. 
$\lvert \cdot \rvert$ and $\lVert \cdot \rVert_2$ denote the absolute value and Euclidean norm, respectively. 
The operators $(\cdot)^T$ and $(\cdot)^{\dagger}$ correspond to the transpose and Hermitian transpose, respectively. $\mathcal{CN}(\delta,\sigma^2)$ represents a complex Gaussian distribution with mean $\delta$ and variance $\sigma^2$. The real part of a complex number $x$ is denoted by $\Re\{x\}$. Finally, $\mathbf{0}_{L \times N}$ denotes the all-zero matrix of dimension $L \times N$.

\begin{figure}[tbp]
\begin{centering}
\includegraphics[scale=.56]{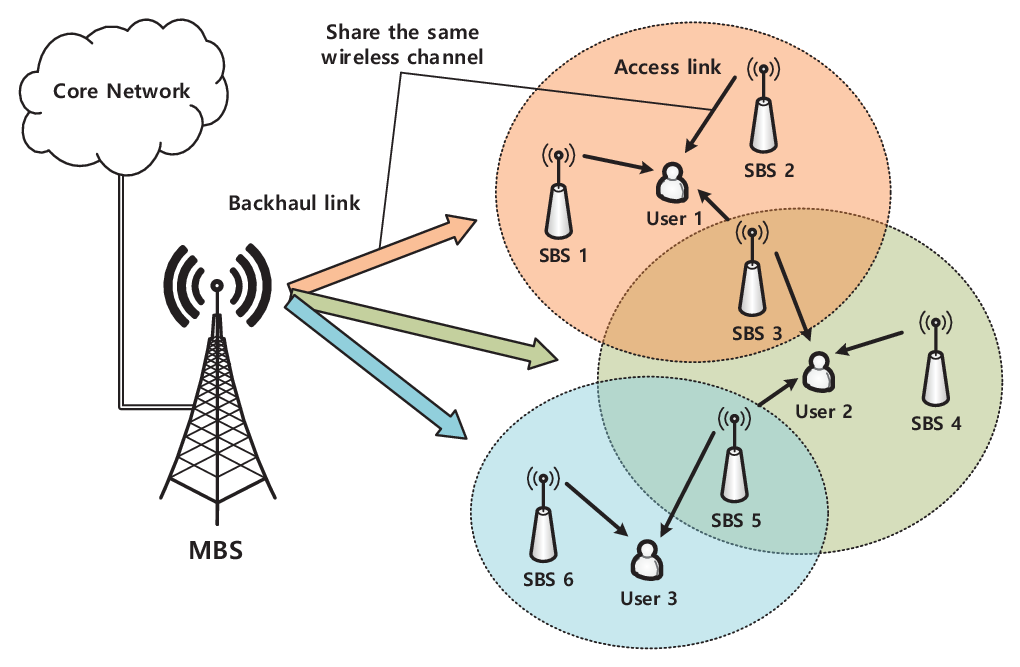}
\caption{\small{User-centric joint access-backhaul transmission in an IBFD self-backhauled wireless network.}}\label{fig:fig_system_model}
\end{centering}
\end{figure}

\section{System Model} \label{sec:System Model}
\subsection{Network Model} \label{sec:Network Model}
Consider the downlink transmission of a wireless network with full-duplex self-backhauling, where $N$ SBSs $\{b_n \mid n \in \mathcal{N} \triangleq \{1,2,\dots,N\}\}$ cooperatively serve $K$ single-antenna users $\{u_k \mid k \in \mathcal{K} \triangleq \{1,2,\dots,K\}\}$ and are all connected to an MBS $b_0$ through in-band wireless backhaul. Each SBS is enabled by a full-duplex radio with $L+1$ antennas: one for receiving at the wireless backhaul (from the MBS to SBSs) and $L$ for transmitting at the access link (from SBSs to users). 
Here, for practical interest, we have adopted the antenna-conserved full-duplex model for each SBS as discussed in \cite{Aryafar_Mobicom12}, where each antenna is associated with a pair of Tx/Rx RF chains and only half the number of RF chains (either Tx or Rx) can be used at each time, as in legacy multi-antenna node. The MBS is equipped with $M$ antennas.
In addition to providing backhaul for the SBSs, we assume that the MBS can also do access to serve its own users directly on other orthogonal resource blocks. This direct access transmission design is a well-known multi-user beamforming problem, thus is not considered in this paper for simplicity.

We consider a user-centric clustering strategy in the access link. Each user $u_k$ is served by a cluster of SBSs cooperatively, denoted as $\mathcal{N}_{k}$. 
An example is shown in Fig. \ref{fig:fig_system_model}, where the serving SBS clusters of the three users are $\mathcal{N}_{1} = \{1,2,3\}$, $\mathcal{N}_{2} = \{3,4,5\}$, and $\mathcal{N}_{3} = \{5,6\}$, respectively.
Let the binary variable $c_{k,n} = 1$ indicate that SBS $b_n$ belongs to the SBS cluster of user $u_k$ and $c_{k,n} = 0$ otherwise. Thus, we have $\mathcal{N}_{k} \triangleq \{n \in \mathcal{N} \mid c_{k,n} = 1\}$. The data intended for user $u_k$ should be fetched at all the SBSs in $\mathcal{N}_{k}$ from the MBS via the backhaul link. Note that the SBS clusters $\{\mathcal{N}_{k}\}_{k=1}^K$ may overlap with each other, which means each SBS may serve multiple users at the same time. We denote $\mathcal{K}_{n} \triangleq \{k \in \mathcal{K} \mid c_{k,n} = 1\}$ as the set of users served by SBS $b_n$.  

In the backhaul link, the MBS adopts a user-centric multicast transmission. Specifically, the MBS treats the message intended for each user $u_k$ as a multicast message and transmits it to the multicast group formed by the SBS cluster $\mathcal{N}_{k}$ associated with this user. Each SBS may receive multiple multicast messages due to the potentially overlapped SBS clusters.
All the messages of the users are superposed and then transmitted simultaneously at the MBS. By carefully designing the multicast beamforming vectors at the MBS, each SBS can decode the set of messages for its served users using successive interference cancellation (SIC)-based receiver.

\begin{table}[!tbp]
\caption{Summary of Notations}  \label{tab:Summary-of-Notations}
\centering
\begin{tabular}{|c|c|} 
\hline
{\bf Notation} & {\bf Description} \\ 
\hline
$\mathcal{N}$ & set of all SBSs \\ 
\hline
$\mathcal{K}$ & set of all users \\ 
\hline
$c_{k,n}$ & binary variable indicating whether or not SBS $b_n$ belongs to the cluster of user $u_k$ \\
\hline
$\mathcal{K}_{n}$ & set of users served by SBS $b_n$ \\ 
\hline
$\mathcal{N}_{k}$ & set of SBSs serving user $u_k$ \\
\hline
$x^{\text{A}}_{k}$ ($x^{\text{B}}_{k}$) & message intended for user $u_k$ in the access (backhaul) link \\ 
\hline
$z_{u_k}$ ($z_{b_n}$) & additive white Gaussian noise at user $u_k$ (SBS $b_n$) \\ 
\hline
$\mathbf{w}_{k,n}$ & beamforming vector at SBS $b_n$ for message $x^{\text{A}}_{k}$ \\
\hline
$\mathbf{v}_{k}$ & beamforming vector at the MBS $b_0$ for message $x^{\text{B}}_{k}$ \\
\hline
$\mathbf{h}_{u_k}^{(b_0)}$ ($\mathbf{h}_{u_k}^{(b_j)}$) & channel vector between the MBS $b_0$ (SBS $b_j$) and user $u_k$ \\
\hline
$\mathbf{h}_{b_n}^{(b_0)}$ ($\mathbf{h}_{b_n}^{(b_j)}$, $j \neq n$) & channel vector between the MBS $b_0$ (SBS $b_j$) and SBS $b_n$ \\
\hline
$\mathbf{h}_{b_n}^{(b_n)}$ & SI channel at SBS $b_n$ \\ 
\hline 
$\beta_{\text{SI}}$ & SI suppression capability of the SBSs \\ 
\hline
$P^{\text{M}}$ ($P^{\text{S}}_n$) & peak transmit power of the MBS (SBS $b_n$) \\ 
\hline
\end{tabular}
\end{table} 

\subsection{Signal Model} \label{sec:Signal Model}
Let $x^{\text{A}}_{k} \in \mathbb{C}$ and $x^{\text{B}}_{k} \in \mathbb{C}$ denote the transmitted signals in the access link and the backhaul link, respectively, that carry the message intended for user $u_k$. All these signals have normalized power of $1$.
Let $\mathbf{w}_{k,n} \in \mathbb{C}^{L \times 1}$ denote the beamforming vector at SBS $b_n$ for message $x^{\text{A}}_{k}$ in the access link and $\mathbf{v}_{k} \in \mathbb{C}^{M \times 1}$ denote the multicast beamforming vector at the MBS for message $x^{\text{B}}_{k}$ in the backhaul link. 
Note that $\mathbf{w}_{k,n} = \mathbf{0}_{L \times 1}$ if $c_{k,n} = 0$, which implies that SBS $b_n$ does not participate in the transmission of message $x_{k}^{\text{A}}$.
The main notations in the system model are summarized in Table \ref{tab:Summary-of-Notations}. 

With full-duplex capability, each SBS can deliver messages to its served users in access while receiving messages from the MBS in backhaul simultaneously using the same frequency band.
\subsubsection{Access Link} \label{sec:Access Link}
The received signal at user $u_k$ is given by
\begin{align} \label{equ:received-signal-user}
y_{u_k} =& \underbrace{\sum_{j \in \mathcal{N}_{k}} \mathbf{h}_{u_k}^{(b_j)\dagger} \mathbf{w}_{k,j} x_{k}^{\text{A}}}_{\text{desired signal}} + \underbrace{\sum_{i=1}^K \mathbf{h}_{u_k}^{(b_0)\dagger} \mathbf{v}_{i} x_{i}^{\text{B}}}_{\text{cross-tier interference}}
+ \underbrace{\sum_{i=1, \, i \neq k}^K \left(\sum_{j \in \mathcal{N}_{i}} \mathbf{h}_{u_k}^{(b_j)\dagger} \mathbf{w}_{i,j} \right) x_{i}^{\text{A}}}_{\text{co-tier interference}} + \underbrace{z_{u_k}}_{\text{noise}},
\end{align}
where $\mathbf{h}_{u_k}^{(b_0)} \in \mathbb{C}^{M\times 1}$ ($\mathbf{h}_{u_k}^{(b_j)} \in \mathbb{C}^{L\times 1}$) is the channel vector between the MBS $b_0$ (SBS $b_j$) and user $u_k$, and $z_{u_k} \sim \mathcal{CN}(0,\sigma_{u_k}^2)$ is the additive white Gaussian noise at user $u_k$.
The channel coefficient $\mathbf{h}_{u_k}^{(b_0)}$ is modeled as $\mathbf{h}_{u_k}^{(b_0)} = \sqrt{\beta_{u_k}^{(b_0)}} \mathbf{g}_{u_k}^{(b_0)}$, where $\beta_{u_k}^{(b_0)} \in \mathbb{R}$ is the large-scale fading coefficient that includes the path loss and shadowing, and $\mathbf{g}_{u_k}^{(b_0)} \sim \mathcal{CN}(\mathbf{0}, \mathbf{1})$ is the small-scale fading coefficient modeled as an independent and identically distributed (i.i.d.) random vector. The channel vector $\mathbf{h}_{u_k}^{(b_j)}$ is modeled in the same manner.
In \eqref{equ:received-signal-user}, the first term is the desired signal transmitted cooperatively by all the SBSs in the cluster $\mathcal{N}_{k}$, the second term is the cross-tier interference transmitted by the MBS over the backhaul link, and the third term presents the co-tier interference transmitted by all the SBSs in the same access link but intended for other users. Note that the cross-tier interference term includes the message for user $u_k$ in the backhaul link, i.e., $x_{k}^{\text{B}}$, but the user does not intend to decode it directly due to the potentially weak channel condition between the MBS and this user. 

Based on \eqref{equ:received-signal-user}, the achievable data rate of user $u_k$ in the access link can be expressed as
\begin{align} \label{equ:access-rate-k}
R_{k}^{\text{A}} = \log \left (1 + \frac {\lvert \sum_{j \in \mathcal{N}_{k}} \mathbf{h}_{u_k}^{(b_j)\dagger} \mathbf{w}_{k,j} \rvert^2} {\Phi_{k} + \sigma_{u_k}^2} \right ),
\end{align}
where $\Phi_{k} = \sum_{i=1}^K \lvert \mathbf{h}_{u_k}^{(b_0)\dagger} \mathbf{v}_{i}\rvert^2 + \sum_{i = 1, \, i \neq k}^K \lvert \sum_{j \in \mathcal{N}_{i}} \mathbf{h}_{u_k}^{(b_j)\dagger} \mathbf{w}_{i,j} \rvert^2$.

\subsubsection{Backhaul Link} \label{sec:Backhaul Link}
The received signal at SBS $b_n$ is given by
\begin{align} \label{equ:received-signal-SBS}
y_{b_n} =& \underbrace{\sum_{k \in \mathcal{K}_n} \mathbf{h}_{b_n}^{(b_0)\dagger} \mathbf{v}_{k} x_{k}^{\text{B}}}_{\text{desired signals}} + \underbrace{\sum_{i = 1, \,i \notin \mathcal{K}_n}^K \mathbf{h}_{b_n}^{(b_0)\dagger} \mathbf{v}_{i} x_{i}^{\text{B}}}_{\text{co-tier interference}} 
+ \underbrace{\sum_{i=1}^K \left (\sum_{j \in \mathcal{N}_{i}} \mathbf{h}_{b_n}^{(b_j)\dagger} \mathbf{w}_{i,j}\right) x_{i}^{\text{A}}}_{\text{cross-tier interference including SI}} + \underbrace{z_{b_n}}_{\text{noise}},
\end{align}
where $\mathbf{h}_{b_n}^{(b_0)} \in \mathbb{C}^{M\times 1}$ ($\mathbf{h}_{b_n}^{(b_j)} \in \mathbb{C}^{L\times 1}$, $j \neq n$) is the channel vector between the MBS $b_0$ (SBS $b_j$) and SBS $b_n$ and $z_{b_n} \sim \mathcal{CN}(0,\sigma_{b_n}^2)$ is the additive white Gaussian noise at SBS $b_n$. 
The channel coefficient $\mathbf{h}_{b_n}^{(b_0)}$ ($\mathbf{h}_{b_n}^{(b_j)}$) is modeled as $\mathbf{h}_{b_n}^{(b_0)} = \sqrt{\beta_{b_n}^{(b_0)}} \mathbf{g}_{b_n}^{(b_0)}$ ($\mathbf{h}_{b_n}^{(b_j)} = \sqrt{\beta_{b_n}^{(b_j)}} \mathbf{g}_{b_n}^{(b_j)}$).
In \eqref{equ:received-signal-SBS}, the first term represents the desired signals intended for the set of users $\mathcal{K}_n$ served by SBS $b_n$, the second term is the co-tier interference caused by the signals intended for other users, and the third term is the cross-tier interference transmitted by all the SBSs over the access link. 
Note that the cross-tier interference term in \eqref{equ:received-signal-SBS} includes the self-interference (SI) term $\sum_{i \in \mathcal{K}_{n}} \mathbf{h}_{b_n}^{(b_n)\dagger} \mathbf{w}_{i,n} x_{i}^{\text{A}}$, where $\mathbf{h}_{b_n}^{(b_n)}$ is the SI channel at SBS $b_n$ and models the residual SI due to imperfect SI cancellation.
The cross-tier interference term also contains the desired signals by users in $\mathcal{K}_n$, i.e., $\{x_{i}^{\text{A}} \mid i \in \mathcal{K}_{n}\}$ but transmitted by other SBSs. In this paper, we assume that the CSI between SBSs can be perfectly estimated and made available at all SBSs. More specifically, each SBS collects its own CSI between other SBSs and sends it to the MBS via the backhaul link. The MBS then collects all the CSI between SBSs and sends it to all the SBSs via multicasting. Since the locations of SBSs are fixed, the CSI between SBSs changes very slowly, this can be done with little signaling overhead. Moreover, each SBS $b_n$ already has the knowledge of its transmitted signals, i.e., $\{x_{i}^{\text{A}} \mid i \in \mathcal{K}_{n}\}$, the cross-tier interference from other SBSs that transmit the same signals can be perfectly canceled at SBS $b_n$.

Based on the above discussion, the achievable rate of decoding the message for user $u_k$ at SBS $b_n$ in the backhaul link is given by
\begin{align} \label{equ:backhaul-rate-k-n}
R_{k,n}^{\text{B}} = \log \left (1 + \frac {\lvert \mathbf{h}_{b_n}^{(b_0)\dagger} \mathbf{v}_{k} \rvert^2} {\Delta_{k,n} + \sigma_{b_n}^2} \right ),
\end{align}
where $\Delta_{k,n} = \sum_{i \in \mathcal{I}_{k,n}} \lvert \mathbf{h}_{b_n}^{(b_0)\dagger} \mathbf{v}_{i} \rvert^2 + \sum_{i = 1, \, i \notin \mathcal{K}_{n}}^{K} \lvert \sum_{j \in \mathcal{N}_{i}} \mathbf{h}_{b_n}^{(b_j)\dagger} \mathbf{w}_{i,j} \rvert^2 + \frac{1}{\beta_{\text{SI}}} \sum_{i \in \mathcal{K}_{n}} \lVert \mathbf{w}_{i,n} \rVert_2^2$ is a term composed of co-tier interference, cross-tier interference, and residual SI. $\mathcal{I}_{k,n} = \{i \in \mathcal{K} \mid H_{i} < H_{k} \text{ or } i \notin \mathcal{K}_{n} \}$ is an index set that is jointly determined by the SIC decoding order and the set of users served by SBS $b_n$. $\beta_{\text{SI}} \geq 1$ is a parameter that reflects the SI suppression capability of the SBSs. The larger this parameter, the better the SI suppression capability. When $\beta_{\text{SI}} = 1$, it means that the SBSs are not able to suppress any SI.

Since the message for user $u_k$ is multicast to the SBS cluster $\mathcal{N}_{k}$, to ensure all the SBSs in $\mathcal{N}_{k}$ can decode the message successfully, the overall transmission rate of the message for user $u_k$ in the backhaul link is limited by the SBS with the worst channel condition, given by
\begin{align} \label{equ:backhaul-rate-k}
R_{k}^{\text{B}} = \min_{n \in \mathcal{N}_{k}} \{ R_{k,n}^{\text{B}} \}.
\end{align}
Considering the backhaul link from the MBS to SBSs and the access link from SBSs to users, the end-to-end achievable rate of user $u_k$ is given by
\begin{align} \label{equ:end-to-end-achievable-rate}
R_{k} = \min \{R_{k}^{\text{A}}, \, R_{k}^{\text{B}}\}.
\end{align}

\section{Joint Access-Backhaul Design with Full CSI} \label{sec:Joint Access-Backhaul Design with Full CSI}
In this section, we consider the joint access-backhaul design when all the CSI is available.
We first provide the problem formulation for the joint design of multicast beamforming in the backhaul link, SBS clustering and beamforming in the access link. We then develop an effective algorithm to solve it via the SLBM approach and the iterative link removal technique.

\subsection{Problem Formulation} 
Our objective is to maximize the end-to-end weighted sum rate of the network through joint design of the backhaul multicast beamforming $\mathbf{v} \triangleq \{\mathbf{v}_{k} \mid k \in \mathcal{K}\}$, the access beamforming $\mathbf{w} \triangleq \{\mathbf{w}_{k,n} \mid k \in \mathcal{K}, \,n \in \mathcal{N}\}$, and the SBS clustering $\mathbf{c} \triangleq \{c_{k,n} \mid k \in \mathcal{K}, \,n \in \mathcal{N}\}$. This problem is mathematically formulated as
\begin{subequations} \label{pro:WSR}
\begin{align} 
\mathcal{P}: \max_{\mathbf{w}, \mathbf{v}, \mathbf{c}} ~& \sum_{k=1}^K \omega_{k} R_{k} \label{obj:WSR} \\ 
\text{s.t.} \quad
& \sum_{k=1}^K \lVert \mathbf{v}_{k} \rVert_2^2 \leq P^{\text{M}}, \label{cons:WSR-MBS-power} \\
& \sum_{k=1}^K \lVert \mathbf{w}_{k,n} \rVert_2^2 \leq P^{\text{S}}_n,~\forall~n \in \mathcal{N}, \label{cons:WSR-SBS-power} \\
& \mathbf{v}_{k} \left(1 - \textstyle \max_{n \in \mathcal{N}} \{c_{k,n}\}\right) = \mathbf{0}_{M \times 1}, ~\forall~k \in \mathcal{K}, \label{cons:WSR-MBS-cluster} \\
& \mathbf{w}_{k,n} (1-c_{k,n}) = \mathbf{0}_{L \times 1},~\forall~k \in \mathcal{K},~n \in \mathcal{N}, \label{cons:WSR-SBS-cluster} 
\end{align} 
\end{subequations} 
where $P^{\text{M}}$ is the peak power of the MBS, $P^{\text{S}}_n$ is the peak power of SBS $b_n$, and $\{\omega_{k}\}_{k=1}^K$ denote the weights accounting for possibly different priorities among all users. For example, to guarantee the fairness among the users, the weights $\{\omega_{k}\}_{k=1}^K$ can be updated according to the proportional fairness criterion.

Problem $\mathcal{P}$ is a non-convex MINLP problem \cite{Burer_nonconvex-MINLP_12}, which is NP-hard in general. Obtaining its optimal solution is challenging due to the non-smoothness and the non-convexity of the rate expression \eqref{equ:end-to-end-achievable-rate} as well as the combinatorial nature of the SBS clustering variable $\mathbf{c}$. Even when the SBS clustering $\mathbf{c}$ is given, problem $\mathcal{P}$ is still non-convex and computationally difficult.

In the following subsections, we first tackle problem $\mathcal{P}$ with a given SBS cluster $\mathbf{c}$, denoted as $\mathcal{P}(\mathbf{c})$. 
We propose to solve it via the SLBM approach with a newly introduced lower-bound approximation for the achievable rate. We then develop a heuristic algorithm to determine the SBS clustering $\mathbf{c}$ based on the iterative link removal technique.

\subsection{Successive Lower-Bound Maximization for $\mathcal{P}(\mathbf{c})$} 
For ease of notation, we rewrite problem $\mathcal{P}$ with a given SBS cluster $\mathbf{c}$, i.e., $\mathcal{P}(\mathbf{c})$ as
\begin{align} \label{pro:WSR-SLBM}
\mathcal{P}(\mathbf{c}): \max_{\mathbf{x}} ~& f(\mathbf{x}) \triangleq \sum_{k=1}^K \omega_{k} R_{k}(\mathbf{x}) \\
\text{s.t.} ~~& \mathbf{x} \in \mathcal{X}_{\mathbf{c}}.  \nonumber
\end{align} 
where $\mathbf{x} \triangleq \{\mathbf{w}, \mathbf{v}\}$ and $\mathcal{X}_{\mathbf{c}}$ is a closed convex set of $\mathbf{x}$ constructed by constraints \eqref{cons:WSR-MBS-power}, \eqref{cons:WSR-SBS-power}, \eqref{cons:WSR-MBS-cluster}, and \eqref{cons:WSR-SBS-cluster} with the given $\mathbf{c}$. 

Problem $\mathcal{P}(\mathbf{c})$ is a non-convex problem with non-smooth and non-convex objective function but convex feasible region. We propose to solve it via the SLBM approach \cite{Razaviyayn_BSUM_2013}.
The main idea of the SLBM approach is to successively maximize a sequence of approximate objective functions \cite{Razaviyayn_BSUM_2013}.
Specifically, starting from a feasible point $\mathbf{x}^{0}$, the algorithm generates a sequence of $\{\mathbf{x}^{t} \}$ according to the update rule
\begin{align} \label{pro:WSR-SLBM-update-rule}
\mathbf{x}^{t} \leftarrow \arg \max_{\mathbf{x} \in \mathcal{X}_{\mathbf{c}}} ~& \hat{f}(\mathbf{x}, \mathbf{x}^{t-1}),
\end{align} 
where $\mathbf{x}^{t-1}$ is the point obtained at the $(t-1)$-th iteration and $\hat{f}(\mathbf{x}, \mathbf{x}^{t-1})$ is an approximation of $f(\mathbf{x})$ at the $t$-th iteration. Typically the approximate function $\hat{f}(\mathbf{x}, \mathbf{x}^{t-1})$ needs to be carefully chosen such that the subproblem \eqref{pro:WSR-SLBM-update-rule} is easy to solve. Moreover, to ensure the convergence of the SLBM algorithm, $\hat{f}(\mathbf{x}, \mathbf{x}^{t-1})$ should be a global lower bound for $f(\mathbf{x})$ and also be tight at $\mathbf{x}^{t-1}$, i.e., $\hat{f}(\mathbf{x}, \mathbf{x}^{t-1}) \leq f(\mathbf{x})$ and $\hat{f}(\mathbf{x}^{t-1}, \mathbf{x}^{t-1}) = f(\mathbf{x}^{t-1})$.

Note that the SLBM approach shares the same idea with many important algorithms such as the successive convex approximation (SCA) \cite{sca1978} and the concave-convex procedure (CCP) \cite{yuille2003cccp}, by successively optimizing an approximate version of the original problem. However, they are different mainly in two aspects \cite{Razaviyayn_BSUM_2013}: 
\begin{itemize}
	\item The SCA or CCP approximates both the objective functions and the feasible sets. On the contrary, the SLBM approximates only the objective function.
	\item The SCA or CCP is applicable only to problems with smooth objectives that are differentiable, while the SLBM is able to handle non-smooth objectives.
\end{itemize}

Recall that $\mathcal{P}(\mathbf{c})$ is a non-convex problem with non-smooth and non-convex objective function but convex feasible set, which is very suitable for the SLBM approach. The SCA or CCP cannot be applied. In the following, we construct an approximate function $\hat{f}(\mathbf{x}, \mathbf{x}^{t-1})$ by introducing a novel concave lower-bound approximation for the non-convex achievable rate expression in the objective function based on SINR convexification, which allows subproblem \eqref{pro:WSR-SLBM-update-rule} to be easily solved with guaranteed convergence. 
For comparison purposes, we first present an existing lower-bound approximation constructed by the WMMSE method \cite{Shi_WMMSE_TSP11}, which has been widely used to deal with varieties of sum rate maximization problems.

\subsubsection{WMMSE-based Concave Lower-Bound Approximation}  \label{sec:WMMSE-based Concave Lower-Bound Approximation} 
We take the non-convex achievable rate expression in the access link \eqref{equ:access-rate-k} with given SBS clustering for example. In the WMMSE method, the relationship between the achievable rate and its mean-square-error (MSE) is established. The key transformation is to rewrite it into the following equivalent form \cite{Dai_Yu_ICCW15}:
\begin{align} \label{equ:WMMSE-RA-k}
R_{k}^{\text{A}}(\mathbf{x}) = \max_{\mathbf{a}_{k}} \left \{ \log(\rho_{k}) - \rho_{k} e_{k} + 1 \right \},
\end{align}
where $\mathbf{a}_{k} \triangleq \{\alpha_{k}, \rho_{k} \}$, $\alpha_{k} \in \mathbb{C}$ is the receive beamformer, $\rho_{k} \in \mathbb{R}$ is a scalar variable associated with the $k$-th user, and $e_{k} \in \mathbb{R}$ is the MSE function defined as
\begin{align}
e_{k} &= \mathbb{E} \left [ \lvert (\alpha_{k})^{\dagger} y_{u_k} - x_{k}^{\text{A}} \rvert^2 \right] \nonumber \\
& = \left (\lvert \textstyle \sum_{j \in \mathcal{N}_{k}} \mathbf{h}_{u_k}^{(b_j)\dagger} \mathbf{w}_{k,j} \rvert^2 + \Phi_{k} + \sigma_{u_k}^2 \right ) \lvert \alpha_{k} \rvert^2 - 2\Re \left \{(\alpha_{k})^{\dagger} ( \textstyle \sum_{j \in \mathcal{N}_{k}} \mathbf{h}_{u_k}^{(b_j)\dagger} \mathbf{w}_{k,j}) \right \} + 1.
\end{align}
Note that in \eqref{equ:WMMSE-RA-k}, with given SBS clustering, the achievable rate $R_{k}^{\text{A}}$ in the access link is a function of the set of beamformers $\mathbf{x}$, denoted as $R_{k}^{\text{A}}(\mathbf{x})$.

The unconstrained optimization problem in the right-hand side of \eqref{equ:WMMSE-RA-k} can be easily solved by using the first-order optimality condition. Its optimal solution $\mathbf{a}_{k}^{*}$ is given by
\begin{align} \label{equ:WMMSE-solutions}
\begin{cases} 
\alpha_{k}^{*} = \left (\lvert \textstyle \sum_{j \in \mathcal{N}_{k}} \mathbf{h}_{u_k}^{(b_j)\dagger} \mathbf{w}_{k,j} \rvert^2 + \Phi_{k} + \sigma_{u_k}^2 \right )^{-1} \left( \textstyle \sum_{j \in \mathcal{N}_{k}} \mathbf{h}_{u_k}^{(b_j)\dagger} \mathbf{w}_{k,j} \right), \\
\rho_{k}^{*} = 1 / e_{k}^{*},
\end{cases}
\end{align}
where $\alpha_{k}^{*}$ is the minimum-mean-square-error (MMSE) receive beamformer and $e_{k}^{*}$ is the resulting MSE. 

Denote
\begin{align} \label{equ:WMMSE-approximation}
{G}_{k}^{\text{A-MMSE}}(\mathbf{x}, \mathbf{a}_{k}) \triangleq \log(\rho_{k}) - \rho_{k} e_{k} + 1.
\end{align}
This function is a global lower bound of $R_{k}^{\text{A}}(\mathbf{x})$, which is derived by the rate-MSE relationship. Then, the WMMSE-based lower-bound approximation for the achievable rate $R_{k}^{\text{A}}(\mathbf{x})$ at a feasible point $\mathbf{x}'$ can be constructed as
\begin{align} \label{equ:WMMSE-based-lower-bound-approximation}
\hat{R}_{k}^{\text{A-MMSE}}(\mathbf{x}, \mathbf{x}') \triangleq {G}_{k}^{\text{A-MMSE}}(\mathbf{x}, \mathbf{a}_{k}(\mathbf{x}')), 
\end{align}
where $\mathbf{a}_{k}(\mathbf{x}')$ is the argument that maximizes ${G}_{k}^{\text{A-MMSE}}(\mathbf{x}, \mathbf{a}_{k})$ with the given $\mathbf{x}'$, i.e.,
\begin{align} 
\mathbf{a}_{k}(\mathbf{x}') = \arg\max_{\mathbf{a}_{k}} {G}_{k}^{\text{A-MMSE}}(\mathbf{x}', \mathbf{a}_{k}).
\end{align}
It is easy to check that
\begin{align} 
R_{k}^{\text{A}}(\mathbf{x}) = \max_{\mathbf{a}_{k}} {G}_{k}^{\text{A-MMSE}}(\mathbf{x}, \mathbf{a}_{k}) \geq {G}_{k}^{\text{A-MMSE}}(\mathbf{x}, \mathbf{a}_{k}(\mathbf{x}')) = \hat{R}_{k}^{\text{A-MMSE}}(\mathbf{x}, \mathbf{x}'),
\end{align}
and
\begin{align} 
R_{k}^{\text{A}}(\mathbf{x}') = \max_{\mathbf{a}_{k}} {G}_{k}^{\text{A-MMSE}}(\mathbf{x}', \mathbf{a}_{k}) = {G}_{k}^{\text{A-MMSE}}(\mathbf{x}', \mathbf{a}_{k}(\mathbf{x}')) = \hat{R}_{k}^{\text{A-MMSE}}(\mathbf{x}', \mathbf{x}').
\end{align}
Thus, $\hat{R}_{k}^{\text{A-MMSE}}(\mathbf{x}, \mathbf{x}')$ is a lower bound of $R_{k}^{\text{A}}(\mathbf{x})$ and it is tight at the point $\mathbf{x}'$. 

\subsubsection{SINR Convexification based Concave Lower-Bound Approximation} 
Since log function is concave and nondecreasing, in order to find a concave lower-bound approximation for the achievable rate, we just need to find a concave lower-bound approximation for the SINR expression in the achievable rates. We refer this approach for convexifying the weighted sum rate maximization problem $\mathcal{P}(\mathbf{c})$ as \textit{SINR convexification}. 
Specifically, for the achievable rate $R_{k}^{\text{A}}(\mathbf{x})$ in \eqref{equ:access-rate-k}, we introduce an auxiliary variable $\gamma_{k}$ such that $\gamma_{k} = \Phi_{k} + \sigma_{u_k}^2$. Then the SINR expression in $R_{k}^{\text{A}}(\mathbf{x})$ can be rewritten into a quadratic-over-linear form $\frac {\lvert \sum_{j \in \mathcal{N}_{k}} \mathbf{h}_{u_k}^{(b_j)\dagger} \mathbf{w}_{k,j} \rvert^2} {\gamma_{k}}$, which is jointly convex in $\mathbf{w}_{k,j} $ and $\gamma_{k} > 0$ \cite[Section 3.1.5]{Boyd_convex_optimization}. 
Taking its first-order Taylor expansion at any feasible point $\{ \mathbf{w}_{k,j}^{'}, \gamma_{k}^{'} \}$, we have
\begin{align} \label{equ:SINR-approximation-access-k-Taylor}
\frac {\lvert \sum_{j \in \mathcal{N}_{k}} \mathbf{h}_{u_k}^{(b_j)\dagger} \mathbf{w}_{k,j} \rvert^2} {\gamma_{k}} \geq \frac{2 \Re \{ (\sum_{j \in \mathcal{N}_{k}} \mathbf{h}_{u_k}^{(b_j)\dagger} \mathbf{w}_{k,j}^{'})^{\dagger} (\sum_{j \in \mathcal{N}_{k}} \mathbf{h}_{u_k}^{(b_j)\dagger} \mathbf{w}_{k,j}) \}}{\gamma_{k}^{'}} - \frac{\lvert \sum_{j \in \mathcal{N}_{k}} \mathbf{h}_{u_k}^{(b_j)\dagger} \mathbf{w}_{k,j}^{'} \rvert^2} {(\gamma_{k}^{'})^2} \gamma_{k},
\end{align}
where the equality holds only when $\mathbf{w}_{k,j} = \mathbf{w}_{k,j}^{'}$ and $\gamma_{k} = \gamma_{k}^{'}$. 

Denote
$u_{k}^{\text{A}} = \frac{\textstyle \sum_{j \in \mathcal{N}_{k}} \mathbf{h}_{u_k}^{(b_j)\dagger} \mathbf{w}_{k,j}^{'}}{\gamma_{k}^{'} }$
and replace $\gamma_{k}$ with $\Phi_{k} + \sigma_{u_k}^2$, then equation \eqref{equ:SINR-approximation-access-k-Taylor} can be rewritten as
\begin{align} \label{equ:SINR-approximation-access-k}
\frac {\lvert \sum_{j \in \mathcal{N}_{k}} \mathbf{h}_{u_k}^{(b_j)\dagger} \mathbf{w}_{k,j} \rvert^2} {\Phi_{k} + \sigma_{u_k}^2} \geq 2\Re\{(u_{k}^{\text{A}})^{\dagger} (\textstyle \sum_{j \in \mathcal{N}_{k}} \mathbf{h}_{u_k}^{(b_j)\dagger} \mathbf{w}_{k,j}) \} - \lvert u_{k}^{\text{A}} \rvert^2 (\Phi_{k} + \sigma_{u_k}^2),
\end{align}
which holds for all $u_{k}^{\text{A}} \in \mathbb{C}$. The equality holds only when $u_{k}^{\text{A}} = u_{k}^{\text{A}*}$, where $u_{k}^{\text{A}*}$ is given by
\begin{align} \label{equ:SINR-approximation-MMSE-receiver}
u_{k}^{\text{A}*} = (\Phi_{k} + \sigma_{u_k}^2)^{-1}(\textstyle \sum_{j \in \mathcal{N}_{k}} \mathbf{h}_{u_k}^{(b_j)\dagger} \mathbf{w}_{k,j}).
\end{align}
It is easy to check that the right-hand side of \eqref{equ:SINR-approximation-access-k} is concave in $\mathbf{x}$ and $u_{k}^{\text{A}}$, respectively. 
It is also seen that $u_{k}^{\text{A}*}$ in \eqref{equ:SINR-approximation-MMSE-receiver} is a scaled version of the MMSE receive beamformer in \eqref{equ:WMMSE-solutions}. By taking $u_{k}^{\text{A}}$ as the receive beamformer, the right-hand side of \eqref{equ:SINR-approximation-access-k} can be viewed as a lower bound of the SINR expression under the receive beamformer $u_{k}^{\text{A}}$, which is tight at the MMSE receive beamformer $u_{k}^{\text{A}*}$.  

Substituting \eqref{equ:SINR-approximation-access-k} into the rate expression $R_{k}^{\text{A}}(\mathbf{x})$ in \eqref{equ:access-rate-k}, we have
\begin{align} 
R_{k}^{\text{A}}(\mathbf{x}) &=\log \left (1 + \frac {\lvert \sum_{j \in \mathcal{N}_{k}} \mathbf{h}_{u_k}^{(b_j)\dagger} \mathbf{w}_{k,j} \rvert^2} {\Phi_{k} + \sigma_{u_k}^2} \right ) \nonumber\\
&\geq \log \left(1 + 2\Re\{(u_{k}^{\text{A}})^{\dagger} (\textstyle \sum_{j \in \mathcal{N}_{k}} \mathbf{h}_{u_k}^{(b_j)\dagger} \mathbf{w}_{k,j}) \} - \lvert u_{k}^{\text{A}} \rvert^2 (\Phi_{k} + \sigma_{u_k}^2)\right),
\end{align}
which holds for all $u_{k}^{\text{A}} \in \mathbb{C}$. Then a global lower bound of $R_{k}^{\text{A}}(\mathbf{x})$ is given by
\begin{align} \label{equ:proposed-approximation-access-k}
{G}_{k}^{\text{A}}(\mathbf{x}, u_{k}^{\text{A}}) \triangleq \log \left(1 + 2\Re\{(u_{k}^{\text{A}})^{\dagger} (\textstyle \sum_{j \in \mathcal{N}_{k}} \mathbf{h}_{u_k}^{(b_j)\dagger} \mathbf{w}_{k,j}) \} - \lvert u_{k}^{\text{A}} \rvert^2 (\Phi_{k} + \sigma_{u_k}^2)\right).
\end{align}
Different from the global lower bound ${G}_{k}^{\text{A-MMSE}}(\mathbf{x}, \mathbf{a}_{k})$ in \eqref{equ:WMMSE-approximation}, which is derived by the rate-MSE relationship, the above global lower bound ${G}_{k}^{\text{A}}(\mathbf{x}, u_{k}^{\text{A}})$ is derived via SINR convexification.
Since the log function is concave and nondecreasing, according to the composition rules in \cite[Section 3.2.4]{Boyd_convex_optimization}, ${G}_{k}^{\text{A}}(\mathbf{x}, u_{k}^{\text{A}})$ is concave in $\mathbf{x}$ and $u_{k}^{\text{A}}$, respectively. There also satisfies that
\begin{align} \label{equ:SUM-approximate-function-access-k}
R_{k}^{\text{A}}(\mathbf{x}) = \max_{ u_{k}^{\text{A}} } {G}_{k}^{\text{A}}(\mathbf{x}, u_{k}^{\text{A}}).
\end{align}
By checking its first-order optimality condition, the optimal solution is given by the MMSE receive beamformer in \eqref{equ:SINR-approximation-MMSE-receiver}.

Then, similar to the WMMSE-based lower-bound approximation in \eqref{equ:WMMSE-based-lower-bound-approximation}, we construct a new concave lower-bound approximation for $R_{k}^{\text{A}}(\mathbf{x})$ based on the relationship \eqref{equ:proposed-approximation-access-k} derived via SINR convexification as
\begin{align} \label{equ:concave-lower-bound-u_k} 
\hat{R}_{k}^{\text{A}}(\mathbf{x}, \mathbf{x}') \triangleq {G}_{k}^{\text{A}}(\mathbf{x}, u_{k}^{\text{A}}(\mathbf{x}')), 
\end{align}
where
\begin{align} 
u_{k}^{\text{A}}(\mathbf{x}') = \arg\max_{u_{k}^{\text{A}}} {G}_{k}^{\text{A}}(\mathbf{x}', u_{k}^{\text{A}}).
\end{align}
Clearly, there holds
\begin{align} 
R_{k}^{\text{A}}(\mathbf{x}) = \max_{u_{k}^{\text{A}}} {G}_{k}^{\text{A}}(\mathbf{x}, u_{k}^{\text{A}}) \geq {G}_{k}^{\text{A}}(\mathbf{x}, u_{k}^{\text{A}}(\mathbf{x}')) = \hat{R}_{k}^{\text{A}}(\mathbf{x}, \mathbf{x}'),
\end{align}
and
\begin{align} 
R_{k}^{\text{A}}(\mathbf{x}') = \max_{u_{k}^{\text{A}}} {G}_{k}^{\text{A}}(\mathbf{x}', u_{k}^{\text{A}}) = {G}_{k}^{\text{A}}(\mathbf{x}', u_{k}^{\text{A}}(\mathbf{x}')) = \hat{R}_{k}^{\text{A}}(\mathbf{x}', \mathbf{x}').
\end{align}
Therefore, $\hat{R}_{k}^{\text{A}}(\mathbf{x}, \mathbf{x}')$ is a lower bound of $R_{k}^{\text{A}}(\mathbf{x})$, which is tight at $\mathbf{x}'$. 

Similar to the WMMSE-based lower-bound approximation, our newly introduced lower-bound approximation via SINR convexification can also be used to handle varieties of weighted sum rate maximization problems, e.g., \cite{Shi_WMMSE_TSP11,Dai_Yu_ICCW15}.
Its superiority over the WMMSE-based approximation will be demonstrated via numerical simulations in Section \ref{sec:Simulation Results}.
% Its superiority over the WMMSE-based approximation will be demonstrated via numerical simulations in Section \ref{sec:Simulation Results}.

Applying the similar method, we can obtain a new concave lower-bound approximation for the non-convex rate expression $R_{k,n}^{\text{B}}$ in \eqref{equ:backhaul-rate-k-n} as
\begin{align} \label{equ:lower-bound-achievable-rate-R-k-b}
\hat{R}_{k,n}^{\text{B}}(\mathbf{x}, \mathbf{x}') \triangleq {G}_{k,n}^{\text{B}}(\mathbf{x}, u_{k,n}^{\text{B}}(\mathbf{x}')),
\end{align}
where the function ${G}_{k,n}^{\text{B}}(\mathbf{x}, \, u_{k,n}^{\text{B}})$ is given by
\begin{align} 
{G}_{k,n}^{\text{B}}(\mathbf{x}, \, u_{k,n}^{\text{B}}) \triangleq \log(1 + 2\Re\{(u_{k,n}^{\text{B}})^{\dagger} (\mathbf{h}_{b_n}^{(b_0)\dagger} \mathbf{v}_{k}) \} - \lvert u_{k,n}^{\text{B}} \rvert^2 (\Delta_{k,n} + \sigma_{b_n}^2)),
\end{align}
and $u_{k,n}^{\text{B}}(\mathbf{x}')$ is
\begin{align}
u_{k,n}^{\text{B}}(\mathbf{x}') = \arg\max_{u_{k,n}^{\text{B}}} {G}_{k,n}^{\text{B}}(\mathbf{x}',u_{k,n}^{\text{B}}).
\end{align}

By utilizing the proposed SINR-convexification based lower-bound approximations in \eqref{equ:concave-lower-bound-u_k} and \eqref{equ:lower-bound-achievable-rate-R-k-b}, we construct a lower bound for the end-to-end achievable rate $R_{k}(\mathbf{x})$ in \eqref{equ:end-to-end-achievable-rate} as
\begin{align} \label{equ:tight-concave-lower-bound-approximation}
\hat{R}_{k}(\mathbf{x}, \mathbf{x}') \triangleq \min \left \{ \hat{R}_{k}^{\text{A}}(\mathbf{x}, \mathbf{x}'), \, \min_{n \in \mathcal{N}_{k}} \left \{ \hat{R}_{k,n}^{\text{B}}(\mathbf{x}, \mathbf{x}') \right \} \right \}.
\end{align}
It is easy to verify that $\hat{R}_{k}(\mathbf{x}, \mathbf{x}') \leq R_{k}(\mathbf{x})$, where the equality holds when $\mathbf{x} = \mathbf{x}'$. Moreover, $\hat{R}_{k}(\mathbf{x}, \mathbf{x}')$ is concave in $\mathbf{x}$, since the pointwise minimum of concave functions is also concave \cite[Section 3.2.3]{Boyd_convex_optimization}. 

Thus, the subproblem \eqref{pro:WSR-SLBM-update-rule} in each iteration of the SLBM algorithm becomes
\begin{align} \label{pro:WSR-SLBM-iteration}
\mathbf{x}^{t} \leftarrow \arg \max_{\mathbf{x} \in \mathcal{X}_{\mathbf{c}}} ~& \hat{f}(\mathbf{x}, \mathbf{x}^{t-1}) \triangleq \sum_{k=1}^K \omega_{k} \hat{R}_{k}(\mathbf{x}, \mathbf{x}^{t-1}).
\end{align} 
Since $\hat{f}(\mathbf{x}, \mathbf{x}^{t-1})$ is a lower-bound approximation of the objective function $f(\mathbf{x})$ and is also tight at $\mathbf{x}^{t-1}$, with a feasible initial point, the iterations of the SLBM algorithm converge to a stationary solution of problem $\mathcal{P}(\mathbf{c})$ \cite{Razaviyayn_BSUM_2013}.
Note that problem \eqref{pro:WSR-SLBM-iteration} is a convex problem with log functions in the objective. It can be approximated by a sequence of second-order cone programming (SOCP) problems \cite{Boyd_SOCP_1998} via the successive approximation method \cite{cvx}. Each SOCP can then be solved with a worst-case computational complexity of $\mathcal{O}(K^4 N (NL+M)^{3})$ via the interior-point methods \cite{Boyd_SOCP_1998} using a general-purpose solver, e.g., SDPT3 in CVX \cite{cvx}.

The details of the SLBM algorithm using the proposed SINR-convexification based lower-bound approximation for solving problem $\mathcal{P}(\mathbf{c})$ are summarized in Alg. \ref{alg:WSR-SLBM}, denoted as SINRC-SLBM.
\begin{algorithm}[!htbp]
\caption{The SINRC-SLBM algorithm for solving problem $\mathcal{P}(\mathbf{c})$} 
\label{alg:WSR-SLBM}
\begin{algorithmic}[0]
\STATE \textbf{Initialization:} Find a feasible point $\mathbf{x}^{0}$ and set $t \leftarrow 1$.
\STATE \textbf{Repeat}
\begin{enumerate}
	\item Update $\mathbf{u}^{t-1} \triangleq \{u_{k}^{\text{A}}(\mathbf{x}^{t-1}), \, u_{k,n}^{\text{B}}(\mathbf{x}^{t-1}),~\forall~k, n \}$ according to the MMSE receive beamformer \eqref{equ:SINR-approximation-MMSE-receiver}.
	\item Update $\mathbf{x}^{t}$ by solving problem \eqref{pro:WSR-SLBM-iteration}.
	\item Set $t \leftarrow t + 1$.
\end{enumerate}
\STATE \textbf{Until} the convergence criterion is met. 
\end{algorithmic}
\end{algorithm}

\subsection{Heuristic Algorithm for $\mathcal{P}$}
Intuitively, larger SBS cluster size can achieve higher access rate, but result in a lower backhaul rate due to multicast transmission, and vice versa. Thus, by controlling the SBS cluster size of each user, we can make a balance between the access rate and the backhaul rate, such that the end-to-end rate of the two-hop transmission is maximized.

With the above observation, we propose a heuristic algorithm based on the iterative link removal technique. Specifically, starting with full cooperation for each user, i.e., $c_{k,n}^{0}=1$, for all $k$ and $n$, we shrink the SBS cluster size by deactivating several (denoted as $J_{\Delta}$) weakest SBS-user links at each iteration and then solve the joint access and backhaul beamforming design problem with the updated SBS cluster. More specifically, at the $t$-th iteration, we solve problem $\mathcal{P}(\mathbf{c}^{t})$ and calculate the transmit power of all the active SBS-user links with $c_{k,n}^{t} = 1$ as $P_{k,n} = \lVert \mathbf{w}_{k,n} \rVert_2^2$. We then sort them in the ascending order and update the SBS cluster $\mathbf{c}^{t+1}$ by deactivating $J_{\Delta}$ SBS-user links with the minimum power.
The iterative procedure terminates when all SBS-links are inactive. Comparing the SBS clusters obtained at each iteration, we then choose the one that achieves the maximum objective value to be the final SBS cluster.
The details of the algorithm are summarized in Alg. \ref{alg:WSR-iterative-link-removal}. 
\begin{algorithm}[!htbp]
\caption{The heuristic algorithm for solving problem $\mathcal{P}$} 
\label{alg:WSR-iterative-link-removal}
\begin{algorithmic}[0]
\STATE \textbf{Initialization:} Initialize $\mathbf{c}^{0}$ with $c_{k,n}^{0} = 1$. Set the iteration index $t \leftarrow 0$ and $J_{\Delta}$. 
\STATE \textbf{While} $ \max\{\mathbf{c}^{t}\} = 1$
\begin{enumerate}
	\item Solve $\mathcal{P}(\mathbf{c}^{t})$ in \eqref{pro:WSR-SLBM} using Alg. \ref{alg:WSR-SLBM}. Denote the objective value as $R(\mathbf{c}^{t})$.
	\item Calculate $P_{k,n} = \| \mathbf{w}_{k,n} \|_2^2$ and sort them in the ascending order.
	\item Update the SBS cluster $\mathbf{c}^{t+1}$ by deactivating $J_{\Delta}$ SBS-user links with the minimum power.
	\item Set $t \leftarrow t + 1$.
\end{enumerate}
\STATE \textbf{End}
\STATE Obtain the final SBS cluster $\mathbf{c}^{i^{*}}$, where $i^{*} = \arg\max_{0 \leq i \leq t} R(\mathbf{c}^{i})$.
\end{algorithmic}
\end{algorithm}

\textit{Complexity:} 
At the $t$-th iteration, we need to solve one problem instance $\mathcal{P}(\mathbf{c}^{t})$ to determine which $J_{\Delta}$ SBS-user links should be removed. Since the maximum number of iterations is $KN/J_{\Delta}$, the overall complexity of the algorithm is $\mathcal{O}(K^5 N^2 (NL+M)^{3}/J_{\Delta})$.

\section{Joint Access-Backhaul Design with Partial CSI}  \label{sec:Joint Access-Backhaul Design with Partial CSI}
The joint access-backhaul design presented in the previous section requires full CSI. However, in an ultra-dense network with a large number of cell sites and users, it is challenging to obtain all the CSI due to the excessive signaling overhead but limited training resources. To handle this challenge and reduce the channel estimation overhead, we consider the joint access-backhaul beamforming design only with partial CSI in this section. We first present the detailed assumption on the CSI availability, then provide the problem formulation for the stochastic beamforming design. We develop a stochastic SLBM algorithm by using the proposed
SINR-convexification based lower-bound approximation and a low-complexity algorithm based on the deterministic lower-bound approximation to solve this problem, respectively.

\subsection{Assumption on CSI Availability and Problem Formulation} 
Generally, the location of the cell sites is fixed and high above the ground. The CSI between the cell sites changes very slowly and can be tracked easily.  
The CSI estimation overhead mainly lies in the channel between the users and the cell sites due to the mobility of the users. Thus, one promising approach to reduce the CSI overhead is to acquire part of the instantaneous CSI between users and cell sites. 
In this work, we only need to acquire the instantaneous information of those links that have main contribution to the performance gain of network cooperation. 
Specifically, the instantaneous CSI from each user to its few nearby SBSs with strong large-scale channel gain and the instantaneous CSI from each user to the MBS are needed. 
The remaining channel links, i.e., the instantaneous channel coefficients of the links between each user and the SBSs that have weak large-scale channel gain can be ignored, since the acquisition of these links will only contribute little to the network performance. 
By using this way, we can reduce the CSI overhead greatly without losing too much of the performance.
Throughout this section, the SBS clustering is assumed to be predetermined based on the large-scale fading, which varies slowly enough. Let $\mathcal{N}_{k}$ denote the predetermined SBS cluster of user $u_k$. 

Recall that we have modeled the channel coefficients $\{\mathbf{h}_{u_k}^{(b_j)} \}$ using the large-scale fading coefficients $\{\beta_{u_k}^{(b_j)} \}$ and the small-scale fading coefficients $\{\mathbf{g}_{u_k}^{(b_j)}\}$ as $\{\mathbf{h}_{u_k}^{(b_j)} = \sqrt{\beta_{u_k}^{(b_j)}} \mathbf{g}_{u_k}^{(b_j)}\}$.
Then, the following two kinds of CSI are assumed to be available: 
\begin{enumerate}
	\item Partial instantaneous CSI: The instantaneous CSI between all the cell sites (including the MBS and all SBSs), i.e., $\{\mathbf{h}_{b_n}^{(b_0)}, \, \mathbf{h}_{b_n}^{(b_j)} \mid j, n \in \mathcal{N},~ j \neq n\}$, the instantaneous CSI between the MBS and all users, i.e., $\{\mathbf{h}_{u_k}^{(b_0)} \mid \, k \in \mathcal{K}\}$, and the instantaneous CSI between each user and its serving SBSs, i.e., $\{\mathbf{h}_{u_k}^{(b_j)} \mid \, j \in \mathcal{N}_{k}, \, k \in \mathcal{K}\}$.
	\item Statistical CSI: The large-scale fading coefficients of the channel links between each user and the SBSs that do not serve it, i.e., $\{\beta_{u_k}^{(b_j)} \mid \, j \notin \mathcal{N}_{k}, \, j \in \mathcal{N}, \, k \in \mathcal{K}\}$, whose instantaneous CSI is unavailable.
\end{enumerate}

Let $\Omega = \{\mathbf{h}_{u_k}^{(b_j)} \mid \, j \notin \mathcal{N}_{k}, \, j \in \mathcal{N}, \, k \in \mathcal{K}\}$ denote the set of unknown instantaneous CSI. 
Since the unknown CSI in $\Omega$ only involves the channel in the access link, we consider the following average achievable rate for the $k$-th user in the access link:
\begin{align} \label{equ:average-achievable-access-rate-k}
\bar{R}_{k}^{\text{A}}(\mathbf{x}) = \mathbb{E}_{\Omega} \left [R_{k}^{\text{A}}(\mathbf{x}; \Omega) \right ],
\end{align}
where $R_{k}^{\text{A}}(\mathbf{x}; \Omega)$ denotes the achievable rate for one realization of $\Omega$ and the expectation is performed over $\Omega$. While for the backhaul link, the achievable rate $R_{k,n}^{\text{B}}(\mathbf{x})$ is deterministic and given by \eqref{equ:backhaul-rate-k-n}.
Therefore, the end-to-end average achievable rate of user $u_k$ is 
\begin{align} \label{equ:end-to-end-average-achievable-rate}
\bar{R}_{k}(\mathbf{x}) = \min  \left \{ \bar{R}_{k}^{\text{A}}(\mathbf{x}), \, \min_{n \in \mathcal{N}_{k}} \left \{ R_{k,n}^{\text{B}}(\mathbf{x}) \right \} \right \}.
\end{align}

We aim to maximize the end-to-end average weighted sum rate of all users via joint access-backhaul beamforming design with partial CSI. This problem is formulated as
\begin{align} \label{pro:WSR-SUM-stochastic}
\mathcal{P}_{\text{S}}(\mathbf{c}): \max_{\mathbf{x} \in \mathcal{X}_{\mathbf{c}}} ~& f(\mathbf{x}) \triangleq \sum_{k=1}^K \omega_{k} \bar{R}_{k}(\mathbf{x}).
\end{align} 
Note that $\mathcal{P}_{\text{S}}(\mathbf{c})$ is a stochastic programming problem with non-smooth and non-convex objective function. We propose to solve it via the stochastic SLBM approach with the proposed SINR-convexification based lower-bound approximation.

\subsection{Stochastic Successive Lower-Bound Maximization for $\mathcal{P}_{\text{S}}(\mathbf{c})$}
A classical method for solving the above stochastic optimization problem $\mathcal{P}_{\text{S}}(\mathbf{c})$ is the sample average approximation (SAA) method \cite{Shapiro_SAA14}. By adopting the SAA method, the stochastic objective function is approximated by an ensemble average and the resulting deterministic optimization problem is then solved by an appropriate numerical algorithm. However, its computation complexity is high, especially for non-convex stochastic optimization problems. 
In this paper, we adopt the stochastic SLBM approach by maximizing an approximate ensemble average at each iteration. To ensure convergence and to facilitate computation, the approximate ensemble average should be a locally tight strongly concave lower bound of the expected objective function \cite{Razaviyayn_SSUM_2016}.
To do so, we generalize the SINR-convexification based lower-bound approximation in \eqref{equ:proposed-approximation-access-k} to a strongly concave version as
\begin{align} \label{equ:ccp-approximate-function-access-k}
\tilde{G}_{k}^{\text{A}}(\mathbf{x}, \mathbf{b}_{k}^{\text{A}}; \Omega) \triangleq \log(1 + 2\Re\{(u_{k}^{\text{A}})^{\dagger} (\textstyle \sum_{j \in \mathcal{N}_{k}} \mathbf{h}_{u_k}^{(b_j)\dagger} \mathbf{w}_{k,j}) \} - \lvert u_{k}^{\text{A}} \rvert^2 (\Phi_{k} + \sigma_{u_k}^2)) \nonumber \\
 - \frac{\gamma}{2} \sum_{n=1}^N \lVert \tilde{\mathbf{w}}_{k,n} - \mathbf{w}_{k,n} \rVert_2^2,
\end{align}
where $\mathbf{b}_{k}^{\text{A}} \triangleq \{u_{k}^{\text{A}}, \, \tilde{\mathbf{w}}_{k,n}\}$ and $\tilde{\mathbf{w}}_{k,n} \in \mathbb{C}^{L \times 1}$ is an auxiliary variable corresponding to $\mathbf{w}_{k,n}$.
Comparing with \eqref{equ:proposed-approximation-access-k}, the quadratic term in \eqref{equ:ccp-approximate-function-access-k} is to make $\tilde{G}_{k}^{\text{A}}$ a strongly concave function in $\mathbf{x}$ with any fixed parameter $\gamma > 0$. 

By checking the first-order optimality condition, it can be verified that
\begin{align} 
R_{k}^{\text{A}}(\mathbf{x}; \Omega) = \max_{ \mathbf{b}_{k}^{\text{A}} } \tilde{G}_{k}^{\text{A}}(\mathbf{x}, \mathbf{b}_{k}^{\text{A}}; \Omega), 
\end{align}
of which the optimal solution is given by
\begin{align} \label{equ:ssum-ccp-update-bA-k}
\begin{cases}
u_{k}^{\text{A}*} = (\Phi_{k} + \sigma_{u_k}^2)^{-1} (\textstyle \sum_{j \in \mathcal{N}_{k}} \mathbf{h}_{u_k}^{(b_j)\dagger} \mathbf{w}_{k,j}), \\
\tilde{\mathbf{w}}_{k,n}^{*} = \mathbf{w}_{k,n}.
\end{cases}
\end{align}

Then, we can construct a strongly concave lower-bound approximation for the rate expression $R_{k}^{\text{A}}(\mathbf{x}; \Omega)$ with a given feasible point $\mathbf{x}'$:
\begin{align} 
\hat{R}_{k}^{\text{A}}(\mathbf{x}, \mathbf{x}'; \Omega) \triangleq \tilde{G}_{k}^{\text{A}}(\mathbf{x}, \mathbf{b}_{k}^{\text{A}}(\mathbf{x}';\Omega); \Omega), 
\end{align}
where
\begin{align} 
\mathbf{b}_{k}^{\text{A}}(\mathbf{x}';\Omega) = \arg\max_{\mathbf{b}_{k}^{\text{A}}} \tilde{G}_{k}^{\text{A}}(\mathbf{x}', \mathbf{b}_{k}^{\text{A}}; \Omega). 
\end{align}

By utilizing the stochastic SLBM approach, at the $t$-th iteration, $\bar{R}_{k}^{\text{A}}(\mathbf{x})$ is approximated by 
\begin{align} 
\bar{R}_{k}^{\text{A}}(\mathbf{x}) = \mathbb{E}_{\Omega} \left [R_{k}^{\text{A}}(\mathbf{x}; \Omega) \right ] \simeq \frac{1}{t} \sum_{i=1}^{t} \hat{R}_{k}^{\text{A}}(\mathbf{x}, \mathbf{x}^{i-1}; \Omega^{i}),
\end{align}
where $\Omega^{i}$ is the $i$-th channel realization and $\mathbf{x}^{i-1}$ is the beamformers obtained from the $(i-1)$-th  iteration. While for the achievable rate $R_{k,n}^{\text{B}}(\mathbf{x})$, it is replaced by its concave lower bound $\hat{R}_{k,n}^{\text{B}}(\mathbf{x}, \mathbf{x}^{t-1})$ derived in \eqref{equ:lower-bound-achievable-rate-R-k-b}.

Therefore, at the $t$-th iteration of the stochastic SLBM algorithm, we solve the following subproblem 
\begin{align} \label{pro:WSR-stochastic-SLBM}
\mathbf{x}^{t} \leftarrow \arg \max_{\mathbf{x} \in \mathcal{X}_{\mathbf{c}}} ~&  \hat{f}(\mathbf{x}, \mathbf{x}^{t-1}) \triangleq \sum_{k=1}^K \omega_{k} \hat{R}_{k}(\mathbf{x}, \mathbf{x}^{t-1}),
\end{align} 
where
\begin{align} 
\hat{R}_{k}(\mathbf{x}, \mathbf{x}^{t-1}) \triangleq \min \left \{ \frac{1}{t} \sum_{i=1}^{t} \hat{R}_{k}^{\text{A}}(\mathbf{x}, \mathbf{x}^{i-1}; \Omega^{i}), \,  \min_{n \in \mathcal{N}_{k}} \left \{\hat{R}_{k,n}^{\text{B}}(\mathbf{x}, \mathbf{x}^{t-1}) \right \} \right \}.
\end{align}
Since $\hat{f}(\mathbf{x}, \mathbf{x}^{t-1})$ is a strongly concave lower-bound approximation of the original objective function $f(\mathbf{x})$ and is a locally tight at $\mathbf{x}^{t-1}$, the iterations generated by the stochastic SLBM algorithm converge to the set of stationary points of $\mathcal{P}_{\text{S}}(\mathbf{c})$ almost surely \cite{Razaviyayn_SSUM_2016}.
The subproblem \eqref{pro:WSR-stochastic-SLBM} in the each iteration of the stochastic SLBM algorithm is convex, which can be approximated by a sequence of SOCPs and solved using the interior-point methods with a worst-case computational complexity of $\mathcal{O}(K^4 N (NL+M)^{3})$.

Finally, we summarize the stochastic SLBM algorithm using the proposed SINR-convexification based lower-bound approximation for solving problem $\mathcal{P}_{\text{S}}(\mathbf{c})$ in Alg. \ref{alg:WSR-stochastic-SLBM}, denoted as SINRC-SSLBM.
\begin{algorithm}[htbp]
\caption{The SINRC-SSLBM algorithm for solving problem $\mathcal{P}_{\text{S}}(\mathbf{c})$} 
\label{alg:WSR-stochastic-SLBM}
\begin{algorithmic}[0]
\STATE \textbf{Initialization:} Randomly generate one feasible point $\mathbf{x}^{0}$ and set the iteration index $t \leftarrow 1$.
\STATE \textbf{Repeat}
\begin{enumerate}
	\item Obtain a new channel realization $\Omega^{t}$.
	\item Update $\mathbf{b}_{k}^{\text{A}}(\mathbf{x}^{t-1};\Omega^{t})$ according to \eqref{equ:ssum-ccp-update-bA-k} for all $k$.
	\item Update $\mathbf{x}^{t}$ by solving problem \eqref{pro:WSR-stochastic-SLBM}.
	\item Set $t \leftarrow t + 1$.
\end{enumerate}
\STATE \textbf{Until} the stopping criterion is met. 
\end{algorithmic}
\end{algorithm}

\subsection{Low-Complexity Algorithm via Deterministic Approximation} \label{sec:Low-Complexity Algorithm via Deterministic Approximation}
In general, solving problem $\mathcal{P}_{\text{S}}(\mathbf{c})$ via the stochastic optimization algorithms often needs a large number of iterations due to the presence of random parameters, which suffers from high computational complexity. 
In this subsection, we derive a deterministic lower-bound approximation for the average achievable rate $\bar{R}_{k}^{\text{A}}(\mathbf{x})$ in \eqref{equ:average-achievable-access-rate-k} and solve the resulting deterministic optimization problem via the SLBM algorithm with low complexity.

Specifically, since $\log(1 + \frac{c}{x})$ is a convex function for any $x > 0$ with a given positive constant $c$, by using Jensen's inequality \cite[Section 3.1.8]{Boyd_convex_optimization} as in \cite{Pan_JSAC17}, a lower bound of the rate $\bar{R}_{k}^{\text{A}}(\mathbf{x})$ can be derived as
\begin{align}
\bar{R}_{k}^{\text{A}}(\mathbf{x}) &\geq \log \left (1 + \frac {\lvert \textstyle \sum_{j \in \mathcal{N}_{k}} \mathbf{h}_{u_k}^{(b_j)\dagger} \mathbf{w}_{k,j} \rvert^2} {\bar{\Phi}_{k} + \sigma_{u_k}^2} \right ),
\end{align}
where
\begin{align}
\bar{\Phi}_{k} = \mathbb{E}_{\Omega} \left [\Phi_{k} \right ] &= \mathbb{E}_{\Omega} \left [\sum_{i=1}^K \lvert \mathbf{h}_{u_k}^{(b_0)\dagger} \mathbf{v}_{i}\rvert^2 + \sum_{i = 1, \, i \neq k}^K \lvert \textstyle \sum_{j \in \mathcal{N}_{i}} \mathbf{h}_{u_k}^{(b_j)\dagger} \mathbf{w}_{i,j} \rvert^2\right ] \nonumber \\
& = \sum_{i=1}^K \lvert \mathbf{h}_{u_k}^{(b_0)\dagger} \mathbf{v}_{i}\rvert^2 + \sum_{i = 1, \, i \neq k}^K \mathbf{w}_{i}^{\dagger} \mathbf{A}_{u_k} \mathbf{w}_{i}.
\end{align}
Here, $\mathbf{w}_{i} \triangleq [\mathbf{w}_{i,1}^{\dagger}, \dots, \mathbf{w}_{i,N}^{\dagger}]^{\dagger}$ is the aggregative access beamforming vector of the $i$-th user. Similarly, define $\mathbf{h}_{u_k} \triangleq [\mathbf{h}_{u_k}^{(b_1)\dagger}, \dots, \mathbf{h}_{u_k}^{(b_N)\dagger}]^{\dagger}$ as the aggregative channel of the $k$-th user in the access link.
Note that since $\mathbf{w}_{i,j} = \mathbf{0}_{L \times 1}$ if $c_{i,j} = 1$, for all $i \in \mathcal{K}$ and $j \in \mathcal{N}$, there holds $\sum_{j \in \mathcal{N}_{i}} \mathbf{h}_{u_k}^{(b_j)\dagger} \mathbf{w}_{i,j} = \mathbf{h}_{u_k}^{\dagger} \mathbf{w}_{i}$. Then $\mathbf{A}_{u_k} = \mathbb{E}_{\Omega} \left [ \mathbf{h}_{u_k} \mathbf{h}_{u_k}^{\dagger}\right ] \in \mathbb{C}^{NL \times NL}$ can be expressed as
\begin{align}
\mathbf{A}_{u_k} = 
	\begin{bmatrix}
	    (\mathbf{A}_{u_k})_{1, 1} & \dots & (\mathbf{A}_{u_k})_{1, N} \\
	    (\mathbf{A}_{u_k})_{2, 1} & \dots & (\mathbf{A}_{u_k})_{2, N} \\
	    \vdots & \ddots & \vdots \\
	    (\mathbf{A}_{u_k})_{N, 1} & \dots & (\mathbf{A}_{u_k})_{N, N}
	\end{bmatrix},
\end{align}
where $(\mathbf{A}_{u_k})_{i,j} \in \mathbb{C}^{L \times L}$ is the block matrix of $\mathbf{A}_{u_k}$ at the $i$-th row and $j$-th column, given by
\begin{align}
(\mathbf{A}_{u_k})_{i,j} = 
\begin{cases}
\mathbf{h}_{u_k}^{(b_i)} \mathbf{h}_{u_k}^{(b_j)\dagger}, & \text{ if } c_{k,i} = c_{k,j} = 1,\\
\beta_{u_k}^{(b_i)} \mathbf{I}_{L \times L}, & \text{ if } i=j,~c_{k,i} = 0,\\
\mathbf{0}_{L \times L}, & \text{ otherwise}.
\end{cases}
\end{align}
It can be easily verified that $\mathbf{A}_{u_k}$ is a positive definite matrix. Thus, we obtain a deterministic lower-bound approximation of $\bar{R}_{k}^{\text{A}}(\mathbf{x})$ as
\begin{align}
\tilde{R}_{k}^{\text{A}}(\mathbf{x}) &= \log \left (1 + \frac {\lvert \textstyle \sum_{j \in \mathcal{N}_{k}} \mathbf{h}_{u_k}^{(b_j)\dagger} \mathbf{w}_{k,j} \rvert^2} {\bar{\Phi}_{k} + \sigma_{u_k}^2} \right ) \nonumber \\
&= \log \left (1 + \frac {\lvert \textstyle \sum_{j \in \mathcal{N}_{k}} \mathbf{h}_{u_k}^{(b_j)\dagger} \mathbf{w}_{k,j} \rvert^2} {\sum_{i=1}^K \lvert \mathbf{h}_{u_k}^{(b_0)\dagger} \mathbf{v}_{i}\rvert^2 + \sum_{i = 1, \, i \neq k}^K \mathbf{w}_{i}^{\dagger} \mathbf{A}_{u_k} \mathbf{w}_{i} + \sigma_{u_k}^2} \right ).
\end{align}

Instead of solving the stochastic optimization problem $\mathcal{P}_{\text{S}}(\mathbf{c})$ in \eqref{pro:WSR-SUM-stochastic}, we now solve the following deterministic optimization problem
\begin{align} \label{pro:WSR-SUM-deterministic}
\mathcal{P}_{\text{D}}(\mathbf{c}): \max_{\mathbf{x} \in \mathcal{X}_{\mathbf{c}}} ~& \sum_{k=1}^K \omega_{k} \tilde{R}_{k}(\mathbf{x}),
\end{align} 
where 
\begin{align}
\tilde{R}_{k}(\mathbf{x}) = \min \left \{ \tilde{R}_{k}^{\text{A}}(\mathbf{x}), \, \min_{n \in \mathcal{N}_{k}} \left \{ R_{k,n}^{\text{B}}(\mathbf{x}) \right \} \right \}.
\end{align} 
Note that similar to problem $\mathcal{P}(\mathbf{c})$ in \eqref{pro:WSR-SLBM}, problem $\mathcal{P}_{\text{D}}(\mathbf{c})$ can be efficiently solved via the SLBM algorithm. A concave lower-bound approximation of the rate $\tilde{R}_{k}^{\text{A}}(\mathbf{x})$ can be obtained as in \eqref{equ:concave-lower-bound-u_k} just by replacing $\Phi_{k}$ with $\bar{\Phi}_{k}$. The overall procedure is similar to Alg. \ref{alg:WSR-SLBM}, which is ignored for simplicity.

\section{Simulation Results} \label{sec:Simulation Results}
\begin{figure}[tbp]
\begin{centering}
\includegraphics[scale=.36]{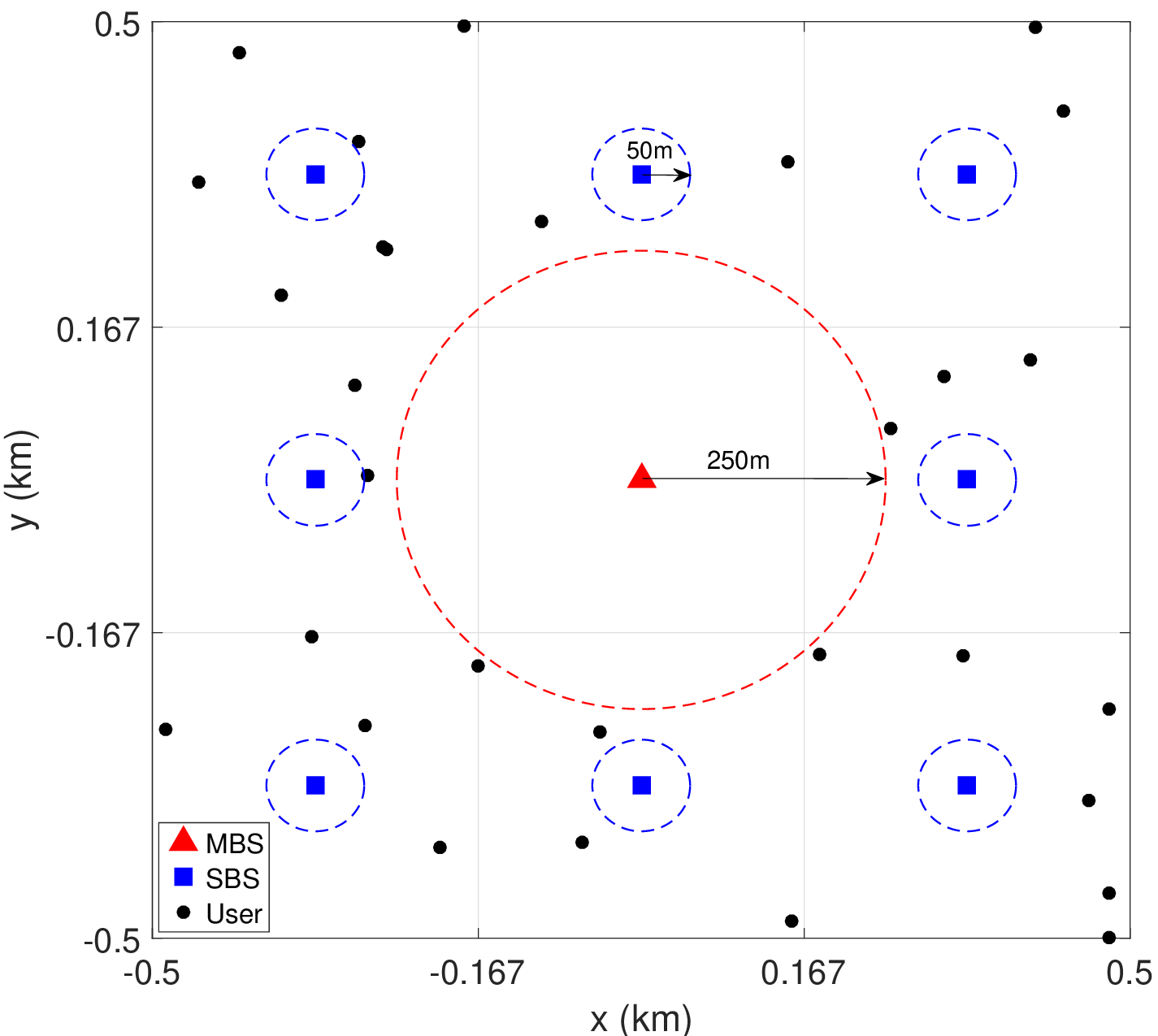}
 \caption{\small{Network topology of a wireless network with one MBS, $8$ SBSs, $30$ mobile users.}}\label{fig:fig_location}
\end{centering}
\end{figure}

In this section, we provide numerical simulations to demonstrate the effectiveness of the proposed algorithms. We consider a wireless network covering a square region of 1 km $\times$ 1 km, as shown in Fig. \ref{fig:fig_location}. The region is divided into a 3-by-3 grid, where one MBS, equipped with $M = 32$ antennas, is located at the center of the region, and $N = 8$ SBSs, each equipped with $L = 2$ antennas for the access link, is located at the center of the rest $8$ small squares. 
The mobile users served by the SBSs are randomly and uniformly distributed within this region, excluding an inner circle of 250 m (50 m) around the MBS (each SBS). In each scheduling interval, $K = 3$ users are scheduled. All the users are scheduled in a round-robin manner. 
The parameter settings are mainly from \cite{3GPP_channel_model_2010} and summarized in Table \ref{tab:parameter-settings}. In particular, the non-line-of-sight (NLOS) path loss models are adopted for the access channel and the line-of-sight (LOS) path loss models are adopted for the backhaul channel.
The iteration of the SINRC-SLBM algorithm in Alg. \ref{alg:WSR-SLBM} stops when the relative increase of the objective value is less than $10^{-3}$ or when a maximum of 30 iterations is reached. For the SINRC-SSLBM algorithm in Alg. \ref{alg:WSR-stochastic-SLBM}, the iteration stops when a maximum of 300 iterations is reached.
For simplicity, the weights $\{\omega_{k}\}_{k=1}^{K}$ in the objective function are set to 1 for all users, and all SBSs have the same peak power, i.e., $P^{\text{S}}_n = P^{\text{S}}$, for all $n \in \mathcal{N}$. The SI suppression capability $\beta_{\text{SI}}$ is set to $110$ dB according to \cite{Pitaval_MWC15_self_backhauling} and \cite{Tabassum_TCOM16_massiveMIMO_FD}.
The convex subproblems of the proposed algorithms, e.g., \eqref{pro:WSR-SLBM-iteration} and \eqref{pro:WSR-stochastic-SLBM} are solved using the CVX package via interior-point solver SDPT3. The experiments are carried out on a Windows x64 machine with 3.3 GHz CPU and 24 GB of RAM.
All the plots are obtained by averaging over 100 channel realizations, if not specified otherwise.

\begin{table}[!tbp]
\caption{Parameter Settings}  \label{tab:parameter-settings}
\centering
\begin{tabular}{l l} 
\hline 
{\bf Parameter} & {\bf Value} \\ 
\hline
Bandwidth & $10$ MHz \\ 
Macro (small) BS antenna gain & $15$ dBi ($5$ dBi) \\ 
Path loss from MBS to user & $128.1 + 37.6 \log_{10}(d)$ dB \\ 
Path loss from SBS to user & $140.7 + 36.7 \log_{10}(d)$ dB \\ 
Path loss from MBS to SBS & $103.4 + 24.2 \log_{10}(d)$ dB \\ 
Path loss from SBS to SBS & $103.8 + 20.9 \log_{10}(d)$ dB \\ 
Macro (small) cell shadowing std. dev. & $8$ dB ($10$ dB) \\ 
Small-scale fading & $\mathcal{CN}(\mathbf{0}, \mathbf{1})$ \\
Noise power $\sigma_{u_k}^2$ ($\sigma_{b_n}^2$) & $-104$ dBm \\ 
\hline
\end{tabular}
\end{table}

\subsection{Convergence Behavior of the Proposed Algorithms} \label{sec:Convergence Behavior of the Proposed Algorithms} 
\begin{figure}[tbp]
\begin{centering}
\includegraphics[scale=.36]{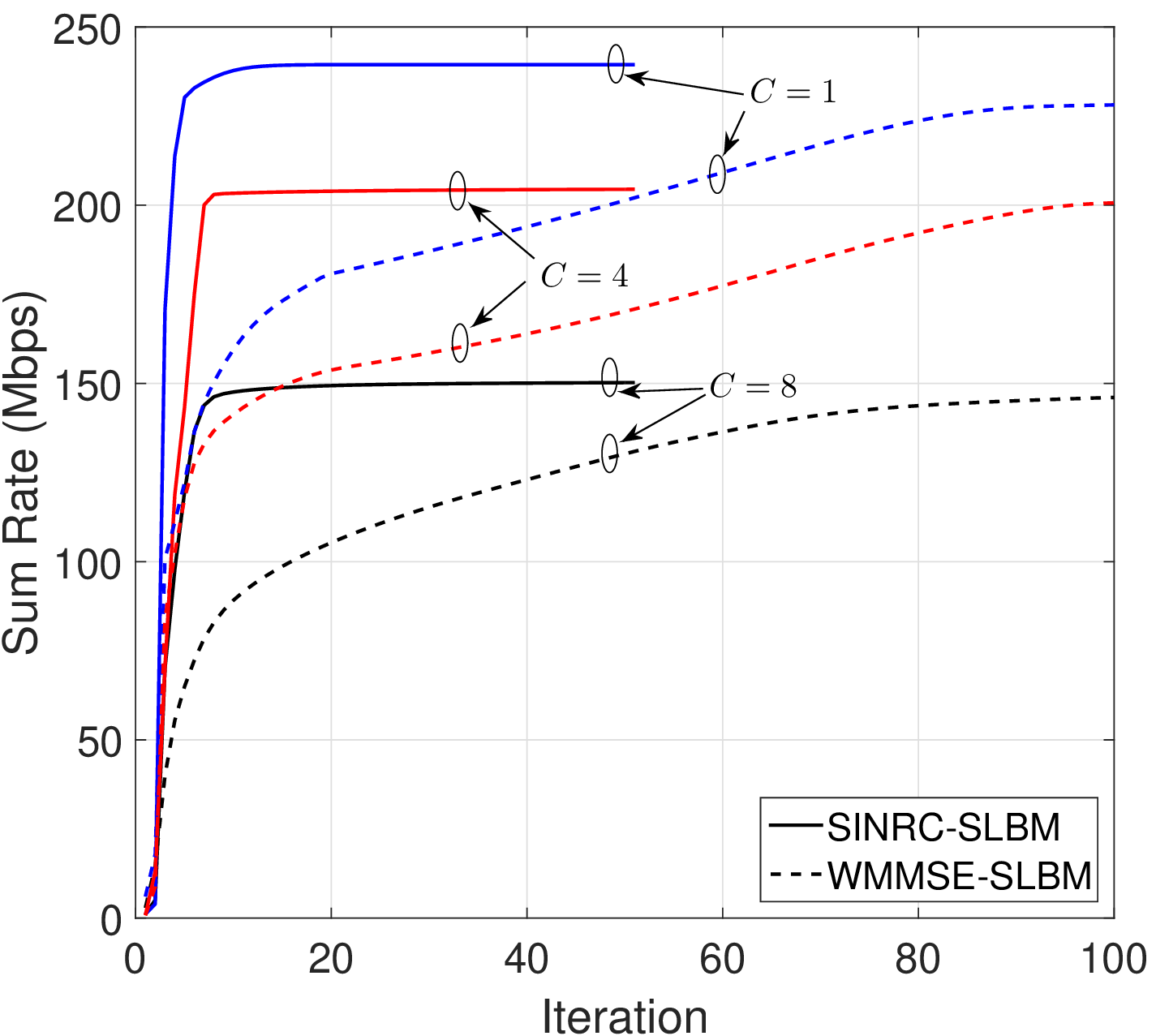}
 \caption{\small{Convergence behavior of the proposed SINRC-SLBM algorithm.}}\label{fig:fig_convergence_SLBM}
\end{centering}
\end{figure}

In this subsection, we first demonstrate the convergence behavior of the SINRC-SLBM algorithm in Alg. \ref{alg:WSR-SLBM} for solving problem $\mathcal{P}(\mathbf{c})$.
We consider a static SBS clustering scheme, where each user $u_k$ is cooperatively served by the cluster of SBSs $\mathcal{N}_{k}$ that have the largest large-scale fading coefficients to the user with a fixed cluster size $\lvert \mathcal{N}_{k} \rvert = C \in \{1, 4, 8\}$. 
The WMMSE algorithm in \cite{Shi_WMMSE_TSP11,Dai_Yu_ICCW15} can be generalized to serve as a benchmark, where $\mathcal{P}(\mathbf{c})$ is solved via the SLBM approach using the WMMSE-based lower-bound approximation \eqref{equ:WMMSE-based-lower-bound-approximation}, denoted as WMMSE-SLBM.
We generate a problem instance and solve it using both the SINRC-SLBM and WMMSE-SLBM algorithms. The maximum transmit power of the MBS and the SBSs are set to $P^{\text{M}} = 40$ dBm and $P^{\text{S}} = 30$ dBm, respectively. 
Fig. \ref{fig:fig_convergence_SLBM} shows the sum rate performance achieved by both algorithms with different cluster sizes. The average per-iteration simulation running time is shown in Table \ref{tab:comparison-of-simulation-time}.
From Fig. \ref{fig:fig_convergence_SLBM}, it is seen that the WMMSE-SLBM algorithm requires about 100 iterations to achieve a good convergence accuracy.
In contrast, the proposed SINRC-SLBM algorithm can converge much faster, all within 10 iterations. It is also seen that the proposed SINRC-SLBM algorithm can converge to the solutions with higher sum rates than those of the WMMSE-SLBM algorithm. 
Although the per-iteration complexity of the proposed SINRC-SLBM algorithm is a little higher than the WMMSE-SLBM algorithm, as shown in Table \ref{tab:comparison-of-simulation-time}, the number of iterations required to converge is 10 times smaller. 
Therefore, the proposed SINRC-SLBM algorithm is more efficient than the WMMSE-SLBM algorithm in terms of both complexity and rate performance, indicating that the proposed SINR-convexification based lower-bound approximation is superior to the WMMSE-based lower-bound approximation.

\begin{table}[tbp]
\caption{Comparison of the average per-iteration simulation time (Seconds)} \label{tab:comparison-of-simulation-time}
\centering
\begin{tabular}{|c|c|c|c|} 
\hline
	 & $C=1$ & $C=4$ & $C=8$ \\ 
\hline
SINRC-SLBM  & 2.08  & 3.51 & 4.12 \\ 
\hline
WMMSE-SLBM & 1.75 & 3.16 & 3.91 \\
\hline
\end{tabular}
\end{table} 

\begin{figure}[tbp]
\begin{centering}
\includegraphics[scale=.36]{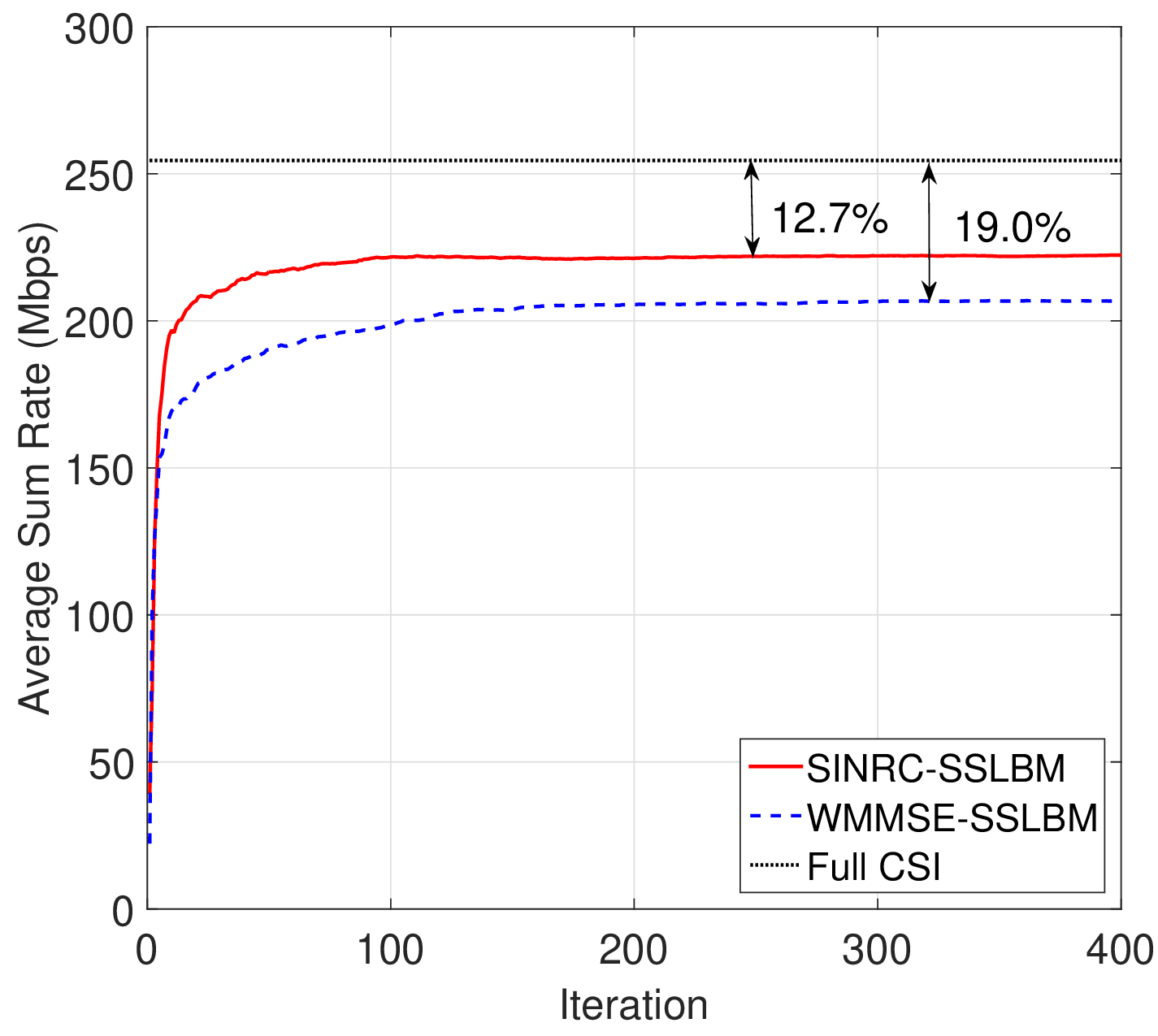}
 \caption{\small{Convergence behavior of the proposed SINRC-SSLBM algorithm.}} \label{fig:fig_convergence_stochastic_SLBM}
\end{centering}
\end{figure}

We also demonstrate the convergence behavior of the SINRC-SSLBM algorithm in Alg. \ref{alg:WSR-stochastic-SLBM} for solving problem $\mathcal{P}_{\text{S}}(\mathbf{c})$.
The stochastic SLBM algorithm proposed in \cite{Razaviyayn_SSUM_2016} using the WMMSE-based lower-bound approximation can serve as a benchmark, denoted as WMMSE-SSLBM.
Fig. \ref{fig:fig_convergence_stochastic_SLBM} shows the average sum rate achieved by both algorithms with cluster size $C = 3$. The average sum rate in each iteration is approximated by the sample mean of 500 independent channel realizations. As can be seen from Fig. \ref{fig:fig_convergence_stochastic_SLBM}, both stochastic algorithms can converge within 200 iterations. However, the proposed SINRC-SSLBM algorithm converges to a solution with better performance than the WMMSE-SSLBM algorithm. 
% This also indicates that the proposed SINR-convexification based lower-bound approximation is superior to the WMMSE-based lower-bound approximation in the stochastic SLBM optimization framework.

\begin{figure}[!tbp]
\begin{centering}
\includegraphics[scale=.36]{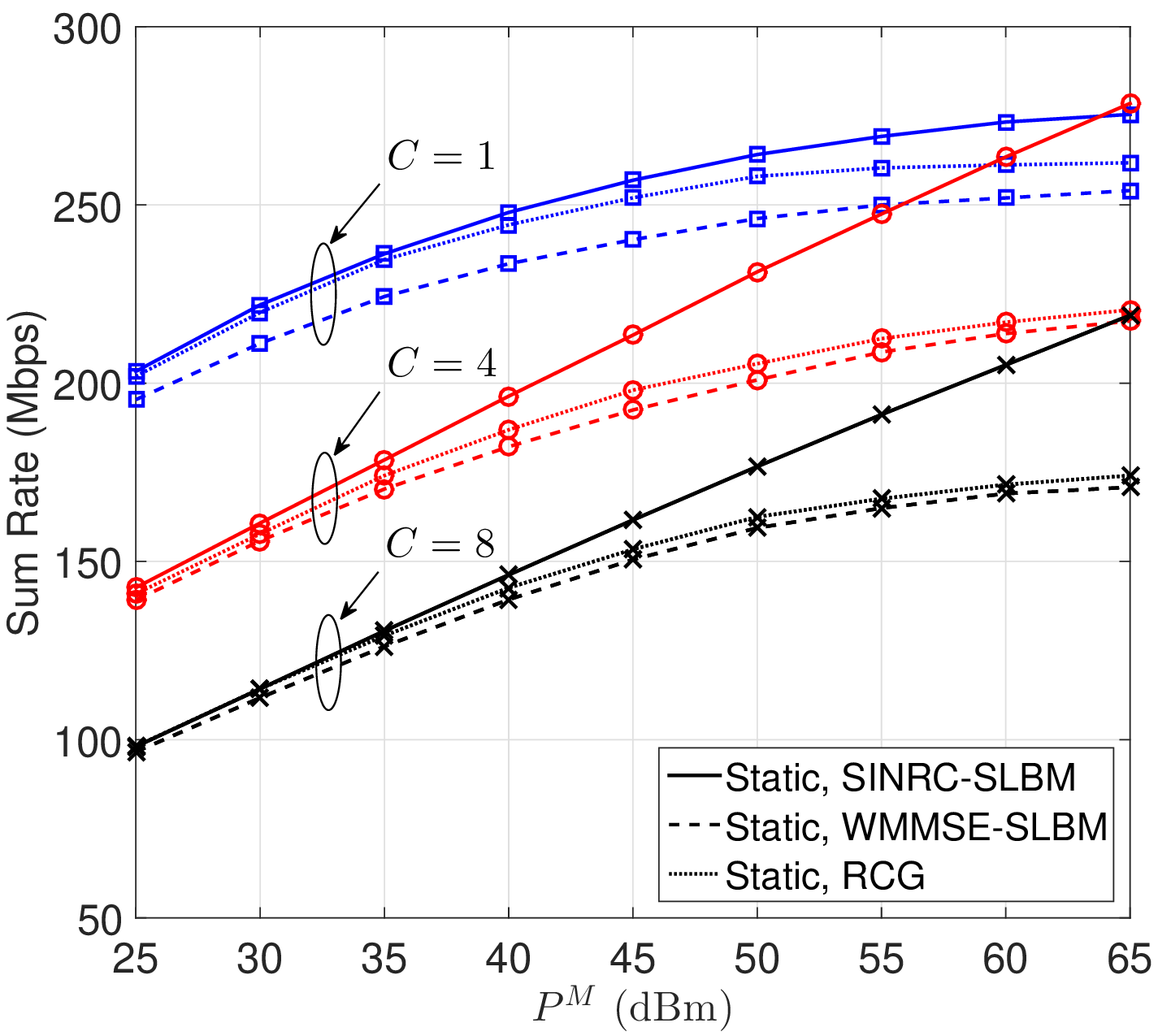}
 \caption{\small{Sum rate versus $P^{\text{M}}$ with $P^{\text{S}} = 30$ dBm.}}\label{fig:fig_benchmark_SLBM}
\end{centering}
\end{figure}

\subsection{Effectiveness of the Proposed Algorithms with Full CSI} \label{sec:Effectiveness of the Proposed Algorithms with Full CSI} 
We first demonstrate the rate performance of the proposed SINRC-SLBM algorithm under the static clustering scheme in Fig. \ref{fig:fig_benchmark_SLBM}. Besides the WMMSE-SLBM, the RCG algorithm proposed in our prior conference paper \cite{Chen_GLOBECOM18} is also considered for comparison.
From Fig. \ref{fig:fig_benchmark_SLBM}, it is clearly seen that the proposed SINRC-SLBM algorithm achieves better performance than the WMMSE-SLBM algorithm as well as the RCG algorithm for all $C \in \{1,4,8\}$. Especially when the peak power of the MBS $P^{\text{M}}$ goes large, the performance gap will increase. This is probably because both the WMMSE-SLBM algorithm and the RCG algorithm are more likely to get stuck in unfavorable local points when $P^{\text{M}}$ is too large, while the proposed SINRC-SLBM algorithm can avoid this high-power issue.

\begin{figure}[tbp]
\begin{centering}
\includegraphics[scale=.36]{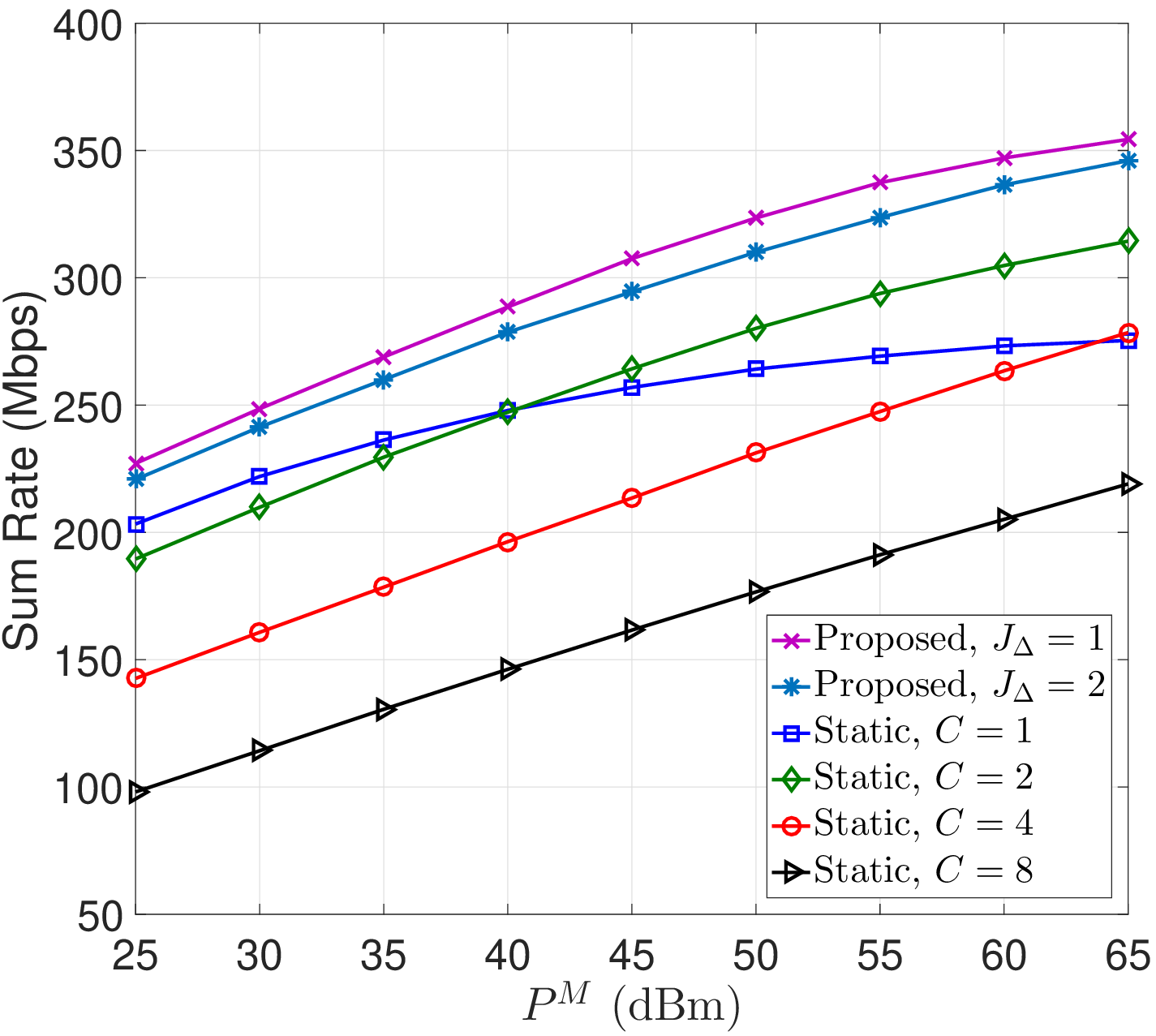}
 \caption{\small{Sum rate versus $P^{\text{M}}$ with $P^{\text{S}} = 30$ dBm.}}\label{fig:fig_benchmark_heuristic}
\end{centering}
\end{figure}

\begin{figure}[!tbp]
\begin{centering}
\includegraphics[scale=.36]{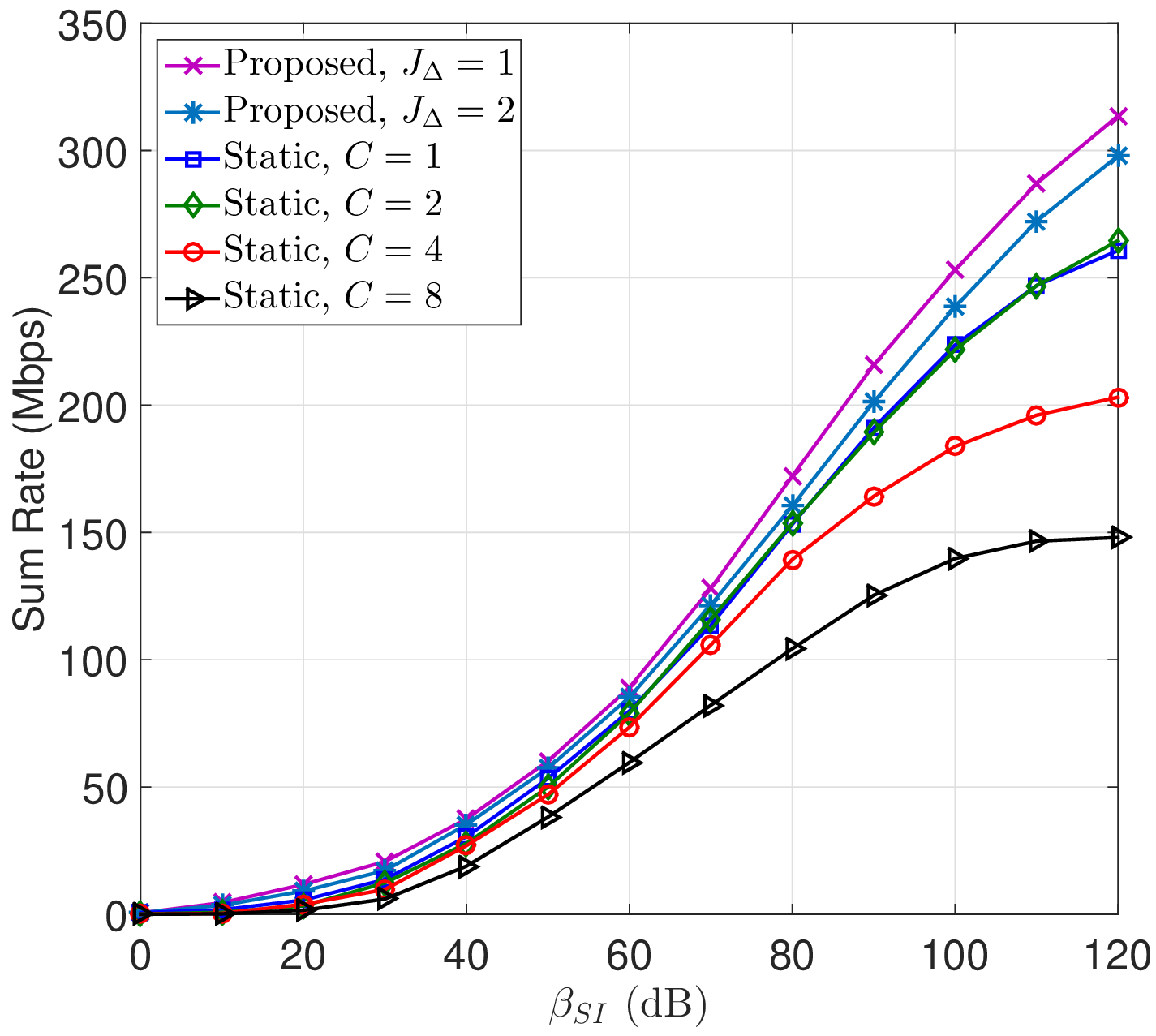}
 \caption{\small{Sum rate versus $\beta_{\text{SI}}$ with $P^{\text{M}} = 40$ dBm and $P^{\text{S}} = 30$ dBm.}}\label{fig:fig_SI_cancellation}
\end{centering}
\end{figure}

\begin{figure}[!tbp]
\begin{centering}
\includegraphics[scale=.38]{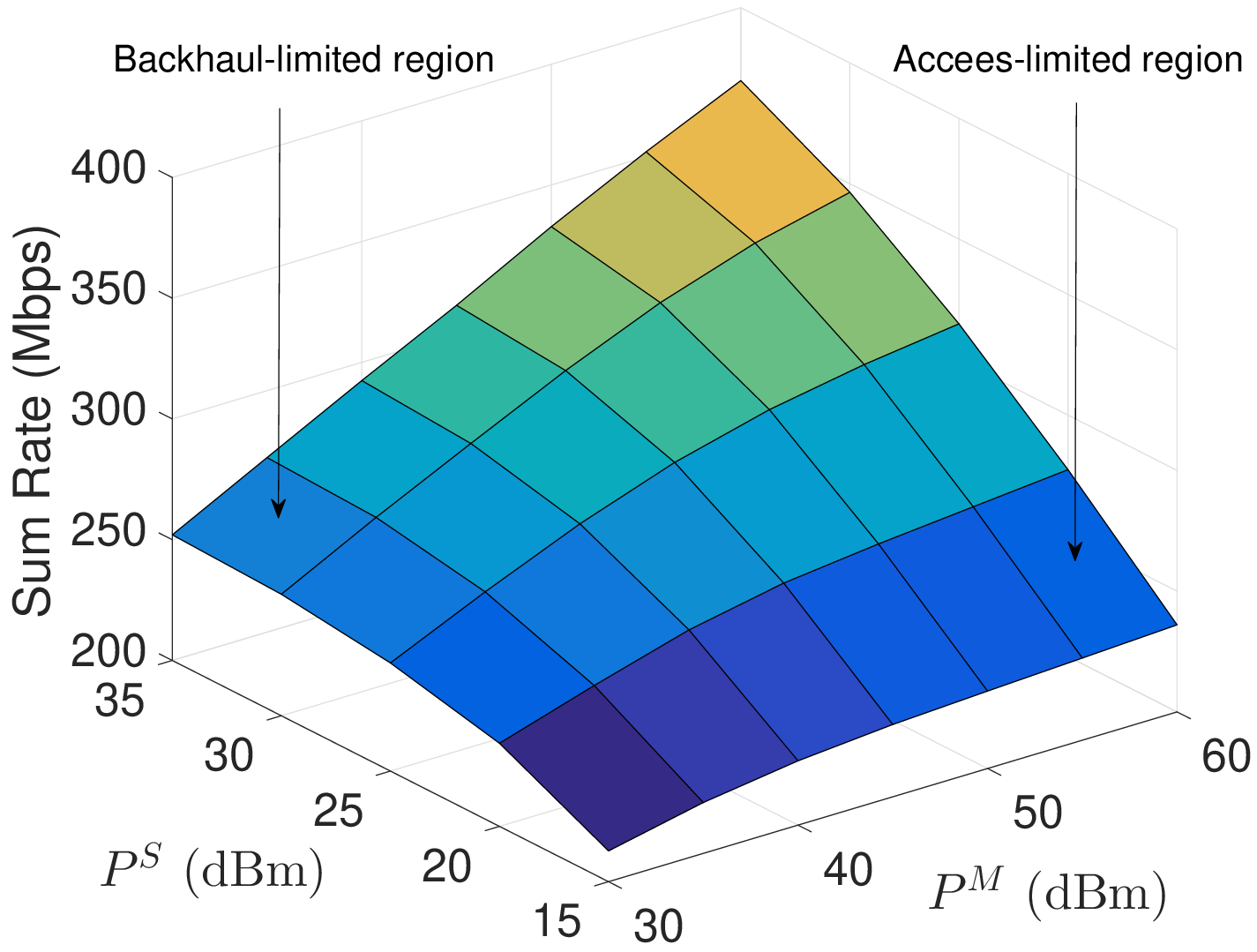}
 \caption{\small{Sum rate versus $P^{\text{M}}$ and $P^{\text{S}}$.}}\label{fig:fig_power_3D}
\end{centering}
\end{figure}

We also demonstrate the performance of the heuristic SBS clustering scheme in Fig. \ref{fig:fig_benchmark_heuristic}. 
For comparison purposes, we consider the static SBS clustering scheme as a benchmark. 
From Fig. \ref{fig:fig_benchmark_heuristic}, it can be seen that the proposed heuristic SBS clustering scheme is superior to the static scheme for all $C \in \{1,2,4,8\}$.
We also see that among all the considered static SBS clustering schemes, the best cluster size $C$ varies at different values of $P^{\text{M}}$. This is due to the fact that higher transmit power in the backhaul link allows more SBSs to cooperate in the access link to increase the end-to-end rate of the two-hop transmission.

The superiority of the proposed heuristic SBS clustering scheme over the static SBS clustering scheme is also demonstrated in Fig. \ref{fig:fig_SI_cancellation} with different SI suppression capability $\beta_{\text{SI}}$. It is observed that the performance of both the heuristic SBS clustering and the static SBS clustering schemes can be greatly improved with the increase of the SI suppression capability. It is also seen that the better the SI suppression capability is, the larger the performance gap between the heuristic SBS clustering scheme and the static SBS clustering scheme will be.

Finally, Fig. \ref{fig:fig_power_3D} illustrates the achievable sum rate performance obtained by the proposed algorithm when both $P^{\text{M}}$ and $P^{\text{S}}$ vary. It is observed that when $P^{\text{S}}$ is small ($\leq 25$ dBm), the sum rate increases slowly when $P^{\text{M}}$ increases. This means that the system performance is limited by the access link. Similarly, when $P^{\text{M}}$ is small ($\leq 40$ dBm), the sum rate increases slowly with $P^{\text{S}}$. This indicates that the system is backhaul-limited. The above observation is expected, since for the two-hop transmission, the end-to-end available rate of a user depends on the minimum rate between the access link and the backhaul link. 

\begin{table}[tbp]
\caption{Sum rate comparison with partial CSI.} \label{tab:Performance-comparison-partial-CSI}
\centering
\begin{tabular}{c|c c c c|c c c c|c c c c} 
\hline
Cluster size & \multicolumn{4}{c|}{$C = 2$} & \multicolumn{4}{c|}{$C = 3$} & \multicolumn{4}{c}{$C = 4$} \\ 
\hline
$P^{\text{M}}$ (dBm) & 30 & 40 & 50 & 60 & 30 & 40 & 50 & 60 & 30 & 40 & 50 & 60  \\ \hline 
SINRC-SSLBM (\%) & 81.2 & 74.4 & 70.0 & 67.5 & 94.3 & 87.3 & 84.1 & 81.2 & 98.5 & 98.1 & 97.9 & 97.0 \\ 
DLB-SLBM (\%)  & 78.4 & 72.4 & 68.0 & 65.5 & 92.4 & 87.2 & 83.7 & 80.6 & 98.1 & 97.7 & 97.6 & 96.8 \\ 
WMMSE-SSLBM (\%)& 78.3 & 69.2 & 63.0 & 60.7 & 86.0 & 81.0 & 74.1 & 68.3 & 89.1 & 87.0 & 82.8 & 79.4 \\
SAA-SLBM (\%)	& 82.4 & 75.8 & 71.9 & 69.3 & 96.1 & 89.5 & 85.7 & 83.2 & 98.5 & 98.2 & 98.1 & 97.8 \\ \hline
Full CSI  (Mbps) & 238.1 & 282.5 & 314.3 & 326.3 & 208.3 & 254.5 & 295.2 & 323.0 & 180.3 & 214.6 & 249.8  & 283.5  \\ \hline
\end{tabular}
\end{table} 

\subsection{Effectiveness of the Proposed Algorithms with Partial CSI} 
\label{sec:Effectiveness of the Proposed Algorithms with Partial CSI} 
Finally, we demonstrate the performance of the two proposed algorithms for partial CSI, namely, the SINRC-SSLBM in Alg. \ref{alg:WSR-stochastic-SLBM} and the low-complexity algorithm via deterministic lower-bound approximation in Section \ref{sec:Low-Complexity Algorithm via Deterministic Approximation}, denoted as DLB-SLBM.
We compare them with the following two benchmark schemes:
\begin{itemize}
	\item \textbf{WMMSE-SSLBM}: The stochastic WMMSE algorithm proposed in \cite{Razaviyayn_SSUM_2016}, which essentially solves problem $\mathcal{P}_{\text{S}}(\mathbf{c})$ via the stochastic SLBM algorithm using the WMMSE-based lower-bound approximation. 
	\item \textbf{SAA-SLBM}: The SAA method, in which the stochastic objective function is approximated by an ensemble average of 200 channel realizations and the resulting deterministic problem is solved by the deterministic SLBM algorithm. 
\end{itemize}

The performance is measured by the percentage of the sum rate achieved by the considered algorithms with partial CSI over the sum rate achieved by the SINRC-SLBM algorithm with full CSI. The results are shown in Table \ref{tab:Performance-comparison-partial-CSI}, where each result is obtained by averaging over 500 independent channel realizations. The static SBS clustering scheme is adopted with a fixed cluster size $C$. The peak power of the SBSs is set to $P^{\text{S}} = 30$ dBm. 
From Table. \ref{tab:Performance-comparison-partial-CSI}, it can be seen that the proposed SINRC-SSLBM algorithm achieves almost the same performance as SAA-SLBM with lower computational complexity. Moreover, it is much better than WMMSE-SSLBM.
It is also seen that the proposed DLB-SLBM algorithm can achieve a good performance that is very close to SAA-SLBM with much lower computational complexity. This clearly demonstrates the effectiveness of the proposed deterministic lower-bound approximation for the average achievable rate. 
From Table. \ref{tab:Performance-comparison-partial-CSI}, we also see that when $C = 4$, with a moderate amount of CSI, the proposed algorithms can maintain good performance that is very close to the full CSI case.
Even when $C = 3$, the proposed algorithms can achieve performance within 80\% $\sim$ 94\% of that with full CSI for varieties of $P^{\text{M}}$.
Therefore, the channel estimation overhead can be significantly reduced with little performance degradation by considering the transmission design with partial CSI. 

\section{Conclusion} \label{sec:Conclusion}
This paper presented a user-centric joint access-backhaul transmission framework for full-duplex self-backhauled wireless networks by exploiting the inter-site cooperation. 
We formulated a weighted sum rate maximization problem for joint design of multicast beamforming in the backhaul link, SBS clustering and beamforming in the access link. 
This problem is solved via the SLBM approach with the newly introduced SINR-convexification based lower-bound approximation and the heuristic iterative link removal technique. 
We also studied the stochastic joint access-backhaul beamforming problem with partial CSI. We developed a stochastic SLBM algorithm and a low-complexity deterministic algorithm to tackle this problem, respectively. 
Simulation results demonstrated the effectiveness of the proposed algorithms for both full CSI and partial CSI scenarios. They also showed that with a moderate amount of CSI, the proposed algorithms can achieve good performance close to that with full CSI. 
By considering the joint transmission design with partial CSI, the proposed algorithms can significantly reduce the CSI overhead without losing too much of the performance.

This initial investigation demonstrated the advantage of the user-centric joint access and backhaul transmission design. There are many interesting directions to pursue in the future. 
For instance, the proposed algorithms are centralized with polynomial complexity, which may be difficult to implement in very large networks. To be more practical, low-complexity or distributed implementation of these algorithms is greatly desired. In addition, a system-level analysis can be carried out using tools from stochastic geometry to give some theoretical results and provide more insights, e.g., the downlink coverage probability and how the system performance varies with the densities of MBSs, SBSs and users.

\bibliographystyle{IEEEtran}
\bibliography{IEEEabrv,IAB}

% Generated by IEEEtran.bst, version: 1.14 (2015/08/26)
\begin{thebibliography}{10}
\providecommand{\url}[1]{#1}
\csname url@samestyle\endcsname
\providecommand{\newblock}{\relax}
\providecommand{\bibinfo}[2]{#2}
\providecommand{\BIBentrySTDinterwordspacing}{\spaceskip=0pt\relax}
\providecommand{\BIBentryALTinterwordstretchfactor}{4}
\providecommand{\BIBentryALTinterwordspacing}{\spaceskip=\fontdimen2\font plus
\BIBentryALTinterwordstretchfactor\fontdimen3\font minus
  \fontdimen4\font\relax}
\providecommand{\BIBforeignlanguage}[2]{{%
\expandafter\ifx\csname l@#1\endcsname\relax
\typeout{** WARNING: IEEEtran.bst: No hyphenation pattern has been}%
\typeout{** loaded for the language `#1'. Using the pattern for}%
\typeout{** the default language instead.}%
\else
\language=\csname l@#1\endcsname
\fi
#2}}
\providecommand{\BIBdecl}{\relax}
\BIBdecl

\bibitem{Chen_GLOBECOM18}
E.~Chen, M.~Tao, and N.~Zhang, ``User-centric joint access-backhaul design for
  full-duplex self-backhauled cooperative networks,'' in \emph{Proc. IEEE
  Global Commun. Conf. (GLOBECOM)}, Dec. 2018, pp. 1--6.

\bibitem{Yuan_5G_ZTE15}
Y.~Yuan and X.~Zhao, ``{5G}: Vision, scenarios and enabling technologies,''
  \emph{ZTE Communications}, vol.~14, no.~S1, pp. 55--60, Dec. 2015.

\bibitem{Ge_wireless_backhaul_network14}
X.~Ge, H.~Cheng, M.~Guizani, and T.~Han, ``{5G} wireless backhaul networks:
  Challenges and research advances,'' \emph{IEEE Network}, vol.~28, no.~6, pp.
  6--11, Nov. 2014.

\bibitem{Pitaval_MWC15_self_backhauling}
R.~A. Pitaval, O.~Tirkkonen, R.~Wichman, K.~Pajukoski, E.~Lahetkangas, and
  E.~Tiirola, ``Full-duplex self-backhauling for small-cell {5G} networks,''
  \emph{IEEE Wireless Commun.}, vol.~22, no.~5, pp. 83--89, Oct. 2015.

\bibitem{Shojaeifard_FD_CRAN_TWC18}
A.~{Shojaeifard}, K.~{Wong}, W.~{Yu}, G.~{Zheng}, and J.~{Tang}, ``Full-duplex
  cloud radio access network: Stochastic design and analysis,'' \emph{IEEE
  Trans. Wireless Commun.}, vol.~17, no.~11, pp. 7190--7207, Nov. 2018.

\bibitem{3GPP_IAB_2018}
{3GPP TR 38.874}, ``{NR; Study on integrated access and backhaul},'' Tech.
  Rep., 2018.

\bibitem{Sharma_TWC17_selfbackhauling}
A.~Sharma, R.~K. Ganti, and J.~K. Milleth, ``Joint backhaul-access analysis of
  full duplex self-backhauling heterogeneous networks,'' \emph{IEEE Trans.
  Wireless Commun.}, vol.~16, no.~3, pp. 1727--1740, Mar. 2017.

\bibitem{Tabassum_TCOM16_massiveMIMO_FD}
H.~Tabassum, A.~H. Sakr, and E.~Hossain, ``Analysis of massive {MIMO}-enabled
  downlink wireless backhauling for full-duplex small cells,'' \emph{IEEE
  Trans. Commun.}, vol.~64, no.~6, pp. 2354--2369, Jun. 2016.

\bibitem{Chen_JSAC16_IBFD_massiveMIMO}
L.~Chen, F.~R. Yu, H.~Ji, B.~Rong, X.~Li, and V.~C.~M. Leung, ``Green
  full-duplex self-backhaul and energy harvesting small cell networks with
  massive {MIMO},'' \emph{IEEE J. Sel. Areas Commun.}, vol.~34, no.~12, pp.
  3709--3724, Dec. 2016.

\bibitem{Debbah_EWC16_IBFD_massiveMIMO}
T.~K. Vu, M.~Bennis, S.~Samarakoon, M.~Debbah, and M.~Latva-aho, ``Joint
  in-band backhauling and interference mitigation in {5G} heterogeneous
  networks,'' in \emph{Proc. 22nd Eur. Wireless Conf.}, May 2016, pp. 1--6.

\bibitem{cooperative_jsac}
D.~Gesbert, S.~Hanly, H.~Huang, S.~Shamai~Shitz, O.~Simeone, and W.~Yu,
  ``Multi-cell {MIMO} cooperative networks: A new look at interference,''
  \emph{IEEE J. Sel. Areas Commun.}, vol.~28, no.~9, pp. 1380--1408, Dec. 2010.

\bibitem{Li_interference_ZTE15}
W.~Li, Y.~Zhang, and L.-K. Huang, ``Interference-cancellation scheme for
  multilayer cellular systems,'' \emph{ZTE Communications}, vol.~13, no.~1, pp.
  43--49, Mar. 2015.

\bibitem{Zhuang_ICC17_CoMP}
H.~Zhuang, J.~Chen, and D.~O. Wu, ``Joint access and backhaul resource
  management for ultra-dense networks,'' in \emph{Proc. IEEE Int. Conf. Commun.
  (ICC)}, May 2017, pp. 1--6.

\bibitem{Hu_TWC17_multicast_clustering}
B.~Hu, C.~Hua, J.~Zhang, C.~Chen, and X.~Guan, ``Joint fronthaul multicast
  beamforming and user-centric clustering in downlink {C-RANs},'' \emph{IEEE
  Trans. Wireless Commun.}, vol.~16, no.~8, pp. 5395--5409, Aug. 2017.

\bibitem{Hua_TWC18}
B.~Hu, C.~Hua, C.~Chen, and X.~Guan, ``Joint beamformer design for wireless
  fronthaul and access links in {C-RANs},'' \emph{IEEE Trans. Wireless
  Commun.}, vol.~17, no.~5, pp. 2869--2881, May 2018.

\bibitem{Ding_NOMA_CM17}
Z.~Ding, Y.~Liu, J.~Choi, Q.~Sun, M.~Elkashlan, C.~L. I, and H.~V. Poor,
  ``Application of non-orthogonal multiple access in {LTE} and {5G} networks,''
  \emph{IEEE Commun. Mag.}, vol.~55, no.~2, pp. 185--191, Feb. 2017.

\bibitem{Kim_IBFD_survey2015}
D.~{Kim}, H.~{Lee}, and D.~{Hong}, ``A survey of in-band full-duplex
  transmission: From the perspective of {PHY} and {MAC} layers,'' \emph{IEEE
  Commun. Surveys Tuts.}, vol.~17, no.~4, pp. 2017--2046, Fourth Quart. 2015.

\bibitem{Krikidis_TWC12}
I.~{Krikidis}, H.~A. {Suraweera}, P.~J. {Smith}, and C.~{Yuen}, ``Full-duplex
  relay selection for amplify-and-forward cooperative networks,'' \emph{IEEE
  Trans. Wireless Commun.}, vol.~11, no.~12, pp. 4381--4393, Dec. 2012.

\bibitem{Ng_TCOM12}
D.~W.~K. {Ng}, E.~S. {Lo}, and R.~{Schober}, ``Dynamic resource allocation in
  {MIMO-OFDMA} systems with full-duplex and hybrid relaying,'' \emph{IEEE
  Trans. Commun.}, vol.~60, no.~5, pp. 1291--1304, May 2012.

\bibitem{Wang_WC15}
G.~Wang, Q.~Liu, R.~He, F.~Gao, and C.~Tellambura, ``Acquisition of channel
  state information in heterogeneous cloud radio access networks: Challenges
  and research directions,'' \emph{IEEE Wireless Commun.}, vol.~22, no.~3, pp.
  100--107, Jun. 2015.

\bibitem{Pan_CM18}
C.~Pan, M.~Elkashlan, J.~Wang, J.~Yuan, and L.~Hanzo, ``User-centric {C-RAN}
  architecture for ultra-dense {5G} networks: Challenges and methodologies,''
  \emph{IEEE Commun. Mag.}, vol.~56, no.~6, pp. 14--20, Jun. 2018.

\bibitem{Shi_ICC14}
Y.~Shi, J.~Zhang, and K.~B. Letaief, ``{CSI} overhead reduction with stochastic
  beamforming for cloud radio access networks,'' in \emph{Proc. IEEE Int. Conf.
  Commun. (ICC)}, Jun. 2014, pp. 5154--5159.

\bibitem{Absil_manifold_optimization}
P.-A. Absil, R.~Mahony, and R.~Sepulchre, \emph{Optimization algorithms on
  matrix manifolds}.\hskip 1em plus 0.5em minus 0.4em\relax Princeton, NJ:
  Princeton University Press, 2008.

\bibitem{Aryafar_Mobicom12}
E.~Aryafar, M.~A. Khojastepour, K.~Sundaresan, S.~Rangarajan, and M.~Chiang,
  ``{MIDU}: Enabling {MIMO} full duplex,'' in \emph{Proc. ACM MobiCom}, 2012,
  pp. 257--268.

\bibitem{Burer_nonconvex-MINLP_12}
S.~Burer and A.~N. Letchford, ``Non-convex mixed-integer nonlinear programming:
  A survey,'' \emph{Surv. Oper. Res. Manage. Sci.}, vol.~17, no.~2, pp.
  97--106, Jul. 2012.

\bibitem{Razaviyayn_BSUM_2013}
M.~Razaviyayn, M.~Hong, and Z.-Q. Luo, ``A unified convergence analysis of
  block successive minimization methods for nonsmooth optimization,''
  \emph{SIAM J. Optim.}, vol.~23, no.~2, pp. 1126--1153, 2013.

\bibitem{sca1978}
B.~R. Marks and G.~P. Wright, ``A general inner approximation algorithm for
  nonconvex mathematical programs,'' \emph{Oper. Res.}, vol.~26, no.~4, pp.
  681--683, Aug. 1978.

\bibitem{yuille2003cccp}
A.~L. Yuille and A.~Rangarajan, ``The concave-convex procedure,'' \emph{Neural
  Comput.}, vol.~15, no.~4, pp. 915--936, 2003.

\bibitem{Shi_WMMSE_TSP11}
Q.~Shi, M.~Razaviyayn, Z.~Luo, and C.~He, ``An iteratively weighted {MMSE}
  approach to distributed sum-utility maximization for a {MIMO} interfering
  broadcast channel,'' \emph{IEEE Trans. Signal Process.}, vol.~59, no.~9, pp.
  4331--4340, Sep. 2011.

\bibitem{Dai_Yu_ICCW15}
B.~Dai and W.~Yu, ``Backhaul-aware multicell beamforming for downlink cloud
  radio access network,'' in \emph{Proc. IEEE Int. Conf. Commun. Workshop
  (ICCW)}, Jun. 2015, pp. 2689--2694.

\bibitem{Boyd_convex_optimization}
S.~Boyd and L.~Vandenberghe, \emph{Convex Optimization}.\hskip 1em plus 0.5em
  minus 0.4em\relax Cambridge, U.K.: Cambridge Univercity Press, 2004.

\bibitem{Boyd_SOCP_1998}
M.~S. Lobo, L.~Vandenberghe, S.~Boyd, and H.~Lebret, ``Applications of
  second-order cone programming,'' \emph{Linear Algebra Appl.}, vol. 284,
  no.~1, pp. 193--228, Nov. 1998.

\bibitem{cvx}
M.~Grant and S.~Boyd, ``{CVX}: Matlab software for disciplined convex
  programming, version 2.1,'' \url{http://cvxr.com/cvx}, Mar. 2014.

\bibitem{Shapiro_SAA14}
A.~Shapiro, D.~Dentcheva, and A.~Ruszczy{\'n}ski, \emph{Lectures on Stochastic
  Programming: Modeling and Theory, Second Edition}.\hskip 1em plus 0.5em minus
  0.4em\relax Philadelphia, PA, USA: Society for Industrial and Applied
  Mathematics, 2014.

\bibitem{Razaviyayn_SSUM_2016}
M.~Razaviyayn, M.~Sanjabi, and Z.-Q. Luo, ``A stochastic successive
  minimization method for nonsmooth nonconvex optimization with applications to
  transceiver design in wireless communication networks,'' \emph{Math.
  Program.}, vol. 157, no.~2, pp. 515--545, Jun. 2016.

\bibitem{Pan_JSAC17}
C.~Pan, H.~Zhu, N.~J. Gomes, and J.~Wang, ``Joint user selection and energy
  minimization for ultra-dense multi-channel {C-RAN} with incomplete {CSI},''
  \emph{IEEE J. Sel. Areas Commun.}, vol.~35, no.~8, pp. 1809--1824, Aug. 2017.

\bibitem{3GPP_channel_model_2010}
{3GPP TR 36.814 V9.0.0}, ``Further advancements for {E-UTRA} physical layer
  aspects,'' Tech. Rep., 2010.

\end{thebibliography}

\end{document}